# Widely distributed exogenic materials of varying compositions and morphologies on asteroid (101955) Bennu

Eri Tatsumi[1,2,3*1], Marcel Popescu[4*1], Humberto Campins[5], Julia de León[1,2], Juan Luis Rizos García[1,2], Javier Licandro[1,2], Amy A. Simon[6], Hannah H. Kaplan[7], Daniella N. DellaGiustina[8], Dathon R. Golish[8], Dante S. Lauretta[8]

1 Instituto de Astrofísica de Canarias (IAC), La Laguna, Spain.
2 Departmento de Astrofísica, Universidad de La Laguna, La Laguna, Spain.
3 Department of Earth and Planetary Science, University of Tokyo, Tokyo, Japan.
4 Astronomical Intsitute of the Romanian Academy, Bucharest, Romania.
5 Department of Physics, University of Central Florida, Orlando, FL, USA.
6 Solar System Exploration Division, NASA Goddard Space Flight Center, Greenbelt, MD, USA.
7 Southwest Research Institute, Boulder, CO, USA.
8 Lunar and Planetary Laboratory, University of Arizona, Tucson, AZ, USA.

**Abstract**

Using the multiband imager MapCam onboard the OSIRIS-REx (Origins, Spectral Interpretation, Resource Identification, and Security–Regolith Explorer) spacecraft, we identified 77 instances of proposed exogenic materials distributed globally on the surface of the B-type asteroid (101955) Bennu. We identified materials as exogenic on the basis of an absorption near 1 µm that is indicative of anhydrous silicates. The exogenic materials are spatially resolved by the telescopic camera PolyCam. All such materials are brighter than their surroundings, and they are expressed in a variety of morphologies: homogeneous, breccia-like, inclusion-like, and others. Inclusion-like features are the most common. Visible spectrophotometry was obtained for 46 of the 77 locations from MapCam images. Principal component analysis indicates at least two trends: (i) mixing of Bennu's average spectrum with a strong 1-µm band absorption, possibly from pyroxene-rich material, and (ii) mixing with a weak 1-µm band absorption. The endmember with a strong 1-µm feature is consistent with Howardite-Eucrite-Diogenite (HED) meteorites, whereas the one showing a weak 1-µm feature may be consistent with HEDs, ordinary chondrites, or carbonaceous chondrites. The variation in the few available near-infrared reflectance spectra strongly suggests varying compositions among the exogenic materials. Thus, Bennu might record the remnants of multiple impacts with different compositions to its parent body, which could have happened in the very early history of the Solar System. Moreover, at least one of the exogenic objects is compositionally different from the exogenic materials found on the similar asteroid (162173) Ryugu, and they suggest different impact tracks.

## 1. Introduction

OSIRIS-REx (Origins, Spectral Interpretation, Resource Identification, and Security–Regolith Explorer) is a NASA mission whose first objective is to return a sample from the near-Earth asteroid (101955) Bennu (Lauretta et al. 2017, 2019, 2021). The spacecraft was launched in September 2016, and the first surface-resolved images were acquired in October 2018.

As a primitive B-type asteroid (Clark et al. 2011), Bennu provides information about the early stages of our Solar System (Lauretta et al. 2015). Bennu, like (162173) Ryugu and (25143) Itokawa, is a rubble pile (e.g., DellaGiustina et al. 2019, Barnouin et al. 2019). Rubble-pile asteroids (reviewed in Walsh 2018) are formed as a result of catastrophic disruption of a parent body and re-accumulation of the fragments by self-gravity. Therefore, reaccumulated rubble piles

---

[1*] These authors contributed equally: E. Tatsumi (etatsumi@iac.es), M. Popescu (popescu.marcel1983@gmail.com).

could include mixtures of materials from both the parent body and its catastrophic impactor, as they did in the case of 2008 TC$_3$ and the related Almahata Sitta meteorites (Jenniskens et al. 2009). Indeed, six unusually bright, basaltic, meter-scale boulders were recently identified on Bennu's dark surface (DellaGiustina et al. 2021), and their close spectral matches to the Howardite-Eucrite-Diogenite (HED) meteorites and Vesta family members led to the hypothesis that they originated from asteroid (4) Vesta. In parallel, bright exogenic anhydrous-silicate–rich materials were found on Ryugu (Tatsumi et al. 2021). The bright boulders on Ryugu are consistent with ordinary chondrite meteorites, based on their albedo and weak or even absent absorption band at 2 μm. These exogenous materials on the surfaces of rubble-pile asteroids could be a key to constraining their specific impact conditions and collisional evolution.

Owing to the small size of Bennu, it is statistically unlikely to have survived more than several hundred million years in the main asteroid belt (e.g. Michel et al. 2009). Thus, Bennu may be the immediate or $n$th-generation progeny of a 100-km-sized body that originally formed in the early Solar System. Bennu's orbit makes it most likely to have originated in the Polana/Eulalia family complex (Campins et al. 2010; Walsh et al. 2013; Bottke et al. 2015); interestingly, this is also the most likely origin of 2008 TC$_3$ (Jenniskens et al. 2009).

Significant compositional changes likely occurred in planetesimals between inside and outside of the "snow line" of the early Solar System. Planetesimals that formed inside the snow line are thought to have been rich in anhydrous silicate materials similar to ordinary chondrites, while those that formed outside would have contained hydrous carbon-rich materials similar to carbonaceous chondrites (Grimm and McSween, 1993). In addition, radial mixing of the primordial planetesimals likely occurred as a result of giant planet migration. Bennu is a carbonaceous asteroid compositionally similar to CM and CI chondrites (Hamilton et al. 2019); hence, the anhydrous silicates detected so far on Bennu are not thought to have originated from Bennu's parent body (DellaGiustina et al. 2021).

In this work, building on the identification of six meter-scale exogenic boulders by DellaGiustina et al. (2021), we study exogenic materials, with diameter of a few tens off centimeter to a few meters, on Bennu's surface between –70° and 70° latitude. First, we survey multi-band images collected by the OSIRIS-REx MapCam to find surface materials with possible 1-μm absorptions as a proxy for mafic minerals such as olivine and pyroxene. Then, we classify them by visible spectrophotometry and morphology. Finally, we discuss the origin of the proposed exogenic materials.

## 2. Detections, spectrophotometry, and spectroscopy
### 2.1. Detection of exogenic materials

MapCam is one of the three cameras in the OSIRIS-REx Camera Suite (OCAMS) (Rizk et al. 2018, Golish et al. 2020). This medium-angle imager is equipped with four chromatic filters (b´: 473 nm, v: 550 nm, w: 698 nm, and x: 847 nm; Golish et al. 2020). The MapCam images used in this work have a pixel scale of ~0.25 m/pixel and were obtained on 26 September 2019 (re-fly of Detailed Survey–Baseball Diamond Flyby 2) (DellaGiustina et al. 2018; Lauretta et al. 2021). A linearity effect must be corrected for measurements close to saturation; hence, for bright objects, we used images with short exposure times. The pixel values need to be below 14000 DN, meaning that we can quantitatively measure the radiance factor up to 7.7% for 10.2-ms exposures and 11.0% for 7.2-ms exposures at the v band (Golish et al. 2020). Each set of color filters (b´, v, w, and x) was co-registered to the v-band image to align the morphological features (DellaGiustina et al. 2020), using a digital shape model of Bennu (Barnouin et al. 2020) produced using stereophotoclinometry (v28) with an 80-cm mean facet size. We normalized each set of color filters by dividing by the v-band image. To expand on the initial discoveries of DellaGiustina et al. (2021), we conducted a search for the absorption in the x band that could be indicative of mafic minerals such as olivine and pyroxene, which are the main composition of HED and ordinary chondrite meteorites.

We used the radiance factor images without photometric correction to avoid introducing spectral variations due to differences in the filters' photometric models. The range of phase angles in the images we used is 8° to 11°. We excluded pixels with incidence and emission angles of 70° or greater, so that color change due to variations in the illumination and observation geometry is negligible. Due to the limitation in illumination angle, the detection will be limited in latitude from -70° to 70°.

To increase signal-to-noise ratio, we binned the images by 2×2 pixels. Taxonomies based on principal component analysis (Tholen 1984, Bus and Binzel, 2002) suggest that the space composed by the linear continuum slope and the 1-µm band absorption can separate stony (S-complex) asteroids from carbonaceous (C-complex) asteroids efficiently. Especially it is important to take into account the slope in the near ultraviolet; S-complex asteroids tend to have strong absorption in this region (Tholen 1984). Thus, we measured the x-band deviation from the continuum, b' to w band linear slope. We used the $D_{xabs}$ index to look for the x-band absorption. $D_{xabs}$ is calculated as:

$$D_{xabs} = 1 - R_x/R_{x0}$$

where $R_x$ is the normalized reflectance at the x band, and $R_{x0}$ is the extrapolated point at the x band based on the spectral slope $\gamma$ defined by the b´ to w band. The slope $\gamma$ was calculated by the least square fitting with following model:

$$R_{band} = 1 + \gamma(\lambda_{band} - 0.55 \text{ µm}) \qquad (band = b´, v, w)$$

where $R_{band}$ is the v-band normalized reflectance and $\lambda_{band}$ is the wavelength at the center of the band filter in microns. Thus, the larger $D_{xabs}$ value indicates deeper x-band absorption. The average $D_{xabs}$ value for Bennu is -0.004±0.055. In our study, material was identified as exogenic if $D_{xabs}$ is larger than 0.15. An additional criterion is that detected objects must appear in multiple image sets. In some cases, the $D_{xabs}$ values are less than 0.15, i.e., the bright boulders are detected as >0.15 in at least one image set. We did not use absolute reflectance in our detection criteria, because absolute reflectance can be easily changed by illumination and observation conditions.

Table 1 shows the 77 instances of proposed exogenic materials found by applying the criteria above. Some of them have $D_{xabs}$ values below 0.15 because this table shows the average value for multiple observations. Their locations are visualized in Fig. 1. We detect all six exogenic boulders reported in DellaGiustina et al. (2021) (EX11, EX15, EX16, EX31, EX36, and EX44). We additionally detect 13 boulders that are also reported in the PYR1b and PHY1b exogenic groups described in Le Corre et al. (2021; see their supplementary materials for which objects overlap between the two studies). That study used the dataset of boulders from DellaGiustina et al. (2020), where the distinct boundaries of boulders bearing a characteristic silicate absorption were identified manually. Our detection method is automated and we do not assume any boulder shape or morphology for identification, which made it possible to identify smaller objects of just a few pixels. In addition, our criteria require absorptions >15%; hence, boulders with weaker absorptions are included in the dataset of Le Corre et al. (2021).

In Fig. A1, we present high-resolution images (<0.05 m/pixel) acquired by PolyCam, OCAMS's narrow-angle panchromatic imager (Rizk et al. 2018, Golish et al. 2020), showing each of these 77 instances in greater detail. There are different morphologies among proposed exogenic materials, such as homogeneous, inclusion-like, breccia-like, etc, which will be discussed in detail in Section 3. The proposed exogenic objects in Table 1 include all morphology groups; boulders including multiple inclusions are indicated by the same ID with sub-numbers.

**Table 1.** Proposed instances of exogenic materials found by applying the criteria described in Section 2.1. See Figure A1 for images of each.

| ID | Latitude (°) | Longitude (°) | b´vw slope (1/um) | $D_{xabs}$ | $R_x/R_w$ |
|---|---|---|---|---|---|
| EX1 | 48.8 | 6.9 | 0.38±0.05 | 0.11±0.01 | 0.97 |
| EX2 | 20.8 | 16.0 | -0.05±0.15 | 0.34±0.05 | 0.76 |
| EX3 | −8.2 | 21.4 | 0.71±0.23 | 0.37±0.04 | 0.75 |
| EX4 | 54.5 | 25.2 | 0.13±0.19 | 0.10±0.05 | 0.95 |
| EX5 | −22.2 | 25.2 | 0.18±0.28 | 0.22±0.10 | 0.88 |
| EX6 | −45.9 | 33.4 | 0.05±0.21 | 0.32±0.04 | 0.80 |
| EX7 | 16.7 | 40.4 | -0.43±0.19 | 0.09±0.08 | 0.97 |
| EX8 | −27.6 | 43.3 | 0.34±0.14 | 0.29±0.02 | 0.86 |
| EX9 | −21.9 | 49.8 | 0.35±0.30 | 0.48±0.05 | 0.65 |
| EX10 | −55.4 | 51.5 | -0.14±0.01 | 0.16±0.03 | 0.95 |
| EX11* | −37.0 | 52.6 | 0.93±0.07 | 0.58±0.02 | 0.53 |
| EX12 | −13.1 | 53.2 | 0.37±0.29 | 0.27±0.17 | 0.82 |
| EX13 | −26.6 | 55.9 | 0.57±0.03 | 0.24±0.02 | 0.90 |
| EX14 | −58.6 | 57.7 | -0.15±0.04 | 0.16±0.02 | 0.92 |
| EX15* | −50.3 | 61.5 | 0.58±0.05 | 0.34±0.01 | 0.76 |
| EX16* | −46.3 | 68.4 | 0.49±0.05 | 0.37±0.01 | 0.73 |
| EX17 | 56.6 | 76.3 | 0.05±0.37 | 0.16±0.08 | 0.93 |
| EX18 | 0.4 | 83.0 | 0.52±0.11 | 0.21±0.04 | 0.92 |
| EX19 | 2.0 | 86.7 | 0.04±0.19 | 0.35±0.03 | 0.77 |
| EX20 | 35.7 | 103.0 | 0.06±0.16 | 0.27±0.08 | 0.83 |
| EX21 | 54.1 | 105.9 | -0.22±0.28 | 0.30±0.16 | 0.82 |
| EX22 | −1.0 | 110.3 | -0.01±0.33 | 0.32±0.09 | 0.76 |
| EX23 | 24.1 | 113.1 | 0.24±0.21 | 0.41±0.04 | 0.68 |
| EX24 | 29.8 | 115.1 | 0.07±0.20 | 0.36±0.03 | 0.74 |
| EX25 | 16.5 | 117.2 | 0.65±0.20 | 0.30±0.03 | 0.85 |
| EX26 | −2.4 | 118.0 | 0.00±0.11 | 0.21±0.02 | 0.85 |
| EX27 | 6.2 | 122.5 | 0.44±0.16 | 0.49±0.02 | 0.64 |
| EX28 | −12.0 | 122.6 | -0.01±0.18 | 0.25±0.13 | 0.82 |
| EX29 | −3.7 | 135.3 | -0.06±0.05 | 0.18±0.01 | 0.93 |
| EX30 | −6.4 | 141.5 | -0.12±0.13 | 0.07±0.01 | 0.95 |
| EX31-1* | −4.1 | 150.5 | 0.20±0.20 | 0.40±0.04 | 0.70 |
| EX31-2* | −3.9 | 151.1 | 0.33±0.09 | 0.39±0.02 | 0.72 |
| EX32 | 15.9 | 157.2 | 0.15±0.19 | 0.22±0.05 | 0.88 |
| EX33 | −13.1 | 158.8 | 0.11±0.17 | 0.23±0.09 | 0.85 |
| EX34 | 23.4 | 159.4 | 0.00±0.17 | 0.20±0.05 | 0.94 |
| EX35 | 21.5 | 163.1 | 0.30±0.02 | 0.14±0.03 | 0.95 |
| EX36* | −43.8 | 166.2 | 0.68±0.11 | 0.24±0.02 | 0.90 |
| EX37 | −16.9 | 170.5 | -0.07±0.28 | 0.31±0.06 | 0.82 |
| EX38 | −13.1 | 177.9 | 0.15±0.20 | 0.20±0.04 | 0.93 |
| EX39 | 13.5 | 180.6 | -0.04±0.08 | 0.29±0.03 | 0.82 |
| EX40 | 22.3 | 181.0 | 0.04±0.19 | 0.16±0.05 | 0.90 |
| EX41 | 4.7 | 183.2 | -0.03±0.57 | 0.34±0.04 | 0.79 |
| EX42 | −31.7 | 186.4 | 0.58±0.17 | 0.26±0.01 | 0.88 |
| EX43 | −20.5 | 189.3 | 0.17±0.35 | 0.22±0.08 | 0.93 |
| EX44* | 12.0 | 189.8 | 0.60±0.07 | 0.51±0.01 | 0.60 |
| EX45 | −4.8 | 190.5 | 0.02±0.15 | 0.33±0.03 | 0.77 |
| EX46 | −43.5 | 198.7 | 0.06±0.22 | 0.28±0.04 | 0.88 |
| EX47 | −22.6 | 199.8 | -0.37±0.50 | 0.23±0.13 | 0.85 |
| EX48 | 10.3 | 204.0 | -0.08±0.21 | 0.28±0.06 | 0.86 |
| EX49 | 2.5 | 213.2 | -0.05±0.17 | 0.26±0.03 | 0.85 |
| EX50 | −16.8 | 213.5 | -0.19±0.22 | 0.14±0.08 | 0.94 |
| EX51 | 31.5 | 218.4 | 0.60±0.06 | 0.24±0.03 | 0.90 |
| EX52 | 30.0 | 220.4 | -0.11±0.11 | 0.24±0.09 | 0.85 |
| EX53-1 | −21.6 | 235.9 | 0.02±0.25 | 0.39±0.05 | 0.73 |
| EX53-2 | −22.4 | 236.4 | -0.36±0.07 | 0.25±0.02 | 0.84 |
| EX54 | −21.2 | 239.2 | -0.05±0.19 | 0.05±0.05 | 0.95 |
| EX55 | −3.5 | 239.6 | -0.09±0.29 | 0.29±0.04 | 0.82 |
| EX56 | 20.1 | 264.4 | -0.07±0.67 | 0.23±0.16 | 0.91 |
| EX57 | −9.8 | 264.6 | 0.23±0.16 | 0.30±0.04 | 0.79 |
| EX58 | 16.9 | 273.1 | 0.45±0.31 | 0.30±0.06 | 0.83 |
| EX59 | −24.5 | 276.5 | 0.33±0.11 | 0.19±0.01 | 0.92 |
| EX60 | −39.1 | 280.5 | -0.07±0.15 | 0.18±0.05 | 0.91 |
| EX61 | 53.4 | 286.1 | 0.37±0.19 | 0.29±0.05 | 0.81 |
| EX62 | 31.7 | 288.9 | 0.34±0.11 | 0.21±0.03 | 0.91 |
| EX63 | 55.1 | 289.9 | 0.36±0.01 | 0.24±0.03 | 0.86 |
| EX64-1 | −43.0 | 291.9 | 0.39±0.27 | 0.28±0.03 | 0.86 |
| EX64-2 | −43.8 | 293.6 | 0.19±0.20 | 0.21±0.04 | 0.94 |
| EX65 | −30.6 | 291.3 | 0.08±0.06 | 0.32±0.01 | 0.76 |
| EX66 | −16.0 | 294.0 | 0.28±0.36 | 0.27±0.11 | 0.81 |
| EX67 | −9.9 | 300.5 | 0.14±0.29 | 0.26±0.07 | 0.85 |
| EX68 | 56.6 | 303.3 | 0.25±0.26 | 0.27±0.02 | 0.83 |
| EX69 | −24.5 | 316.9 | 0.34±0.04 | 0.30±0.06 | 0.78 |
| EX70 | −30.2 | 319.2 | 0.03±0.10 | 0.26±0.09 | 0.81 |
| EX71 | 58.4 | 319.3 | 0.23±0.03 | 0.15±0.01 | 0.93 |
| EX72 | 19.2 | 342.6 | 0.11±0.53 | 0.43±0.07 | 0.70 |
| EX73 | 24.2 | 342.6 | 0.18±0.44 | 0.37±0.09 | 0.76 |
| EX74 | 62.8 | 346.1 | 0.14±0.05 | 0.15±0.04 | 0.93 |

* Boulders previously identified in DellaGiustina et al. (2020a).

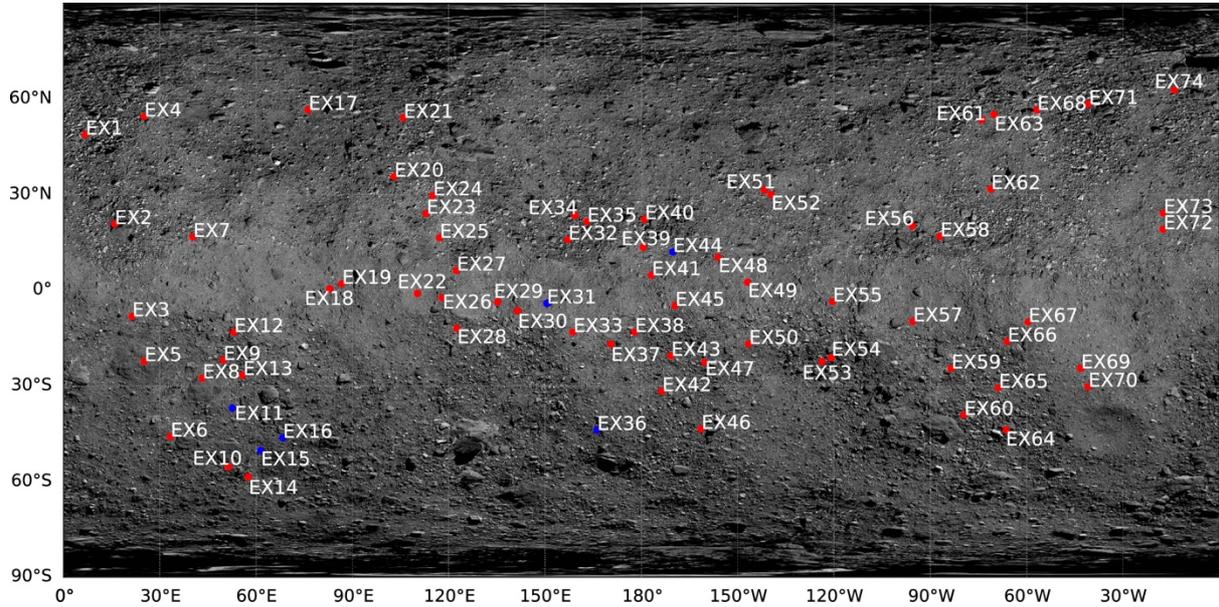

**Figure 1**. Locations of proposed exogenic materials, overlaid on a global mosaic basemap of asteroid Bennu (Bennett et al. 2020). Light blue points are boulders found previously by DellaGiustina et al. (2021), and red points are proposed exogenic materials found in this work that meet our criteria described in Section 2.1.

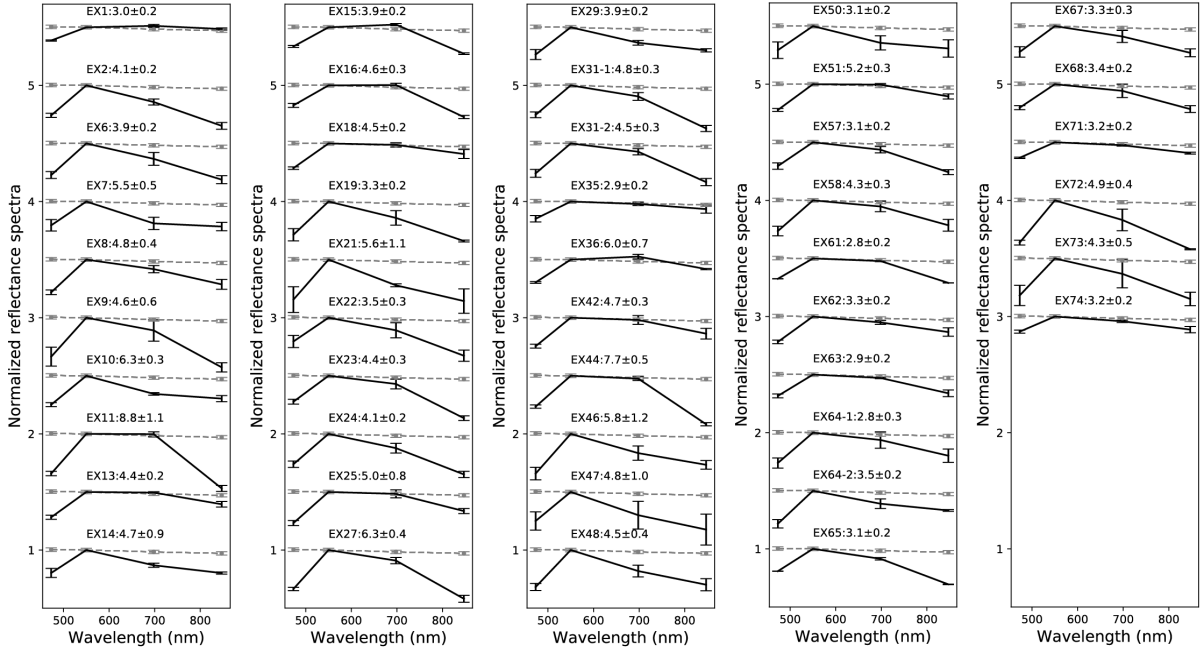

**Figure 2.** Visible spectrophotometry from MapCam images normalized at 550 nm (v-filter) of 46 instances of proposed exogenic material. The reflectance factors (in %) at (*i,e*, $\alpha$) = (30°,0°,30°) are shown for each spectrum. Gray dashed lines indicate the average spectrum of Bennu. Error bars represent standard deviation among multiple observational sets.

## 2.2. MapCam spectrophotometry

We conducted spectrophotometric measurements for candidate areas of exogenic material larger than 0.4 m, a size comparable to the MapCam pixel scale; areas smaller than 0.4 m lack the resolution to measure the spectrophotometric curves accurately. With this cut-off, we could measure the visible spectrophotometry of 46 of the 77 sites. A photometric correction was applied to our candidate areas to measure the absolute reflectance value at v band. The radiance factor images were photometrically corrected using the ROLO (Robotic Lunar Observatory,) model (Buratti et al. 2011), where the parameters ($C_n$, $A_m$, $n$ = 1 and 2, $m$ = 1, 2, 3, and 4) for Bennu were proposed by Golish et al. (2021) to the standard illumination and observing geometry: incidence ($i$ = 30°), emission ($e$ = 0°), and phase ($\alpha$ = 30°) angles. In this model:

Reflectance=A($\alpha$)·d($\alpha,i,e$)

A($\alpha$)=$C_0$ exp(-$C_1\alpha$)+$A_0$+$A_1\alpha$+$A_2\alpha^2$+$A_3\alpha^3$+$A_4\alpha^4$

d($e, i$)=cos($i$)/(cos($i$)+cos($e$))

The necessary phase, emission, and incidence backplanes for making photometric corrections are obtained using ray tracing techniques within the Integrated Software for Imagers and Spectrometers (ISIS3) software (Keszthelyi et al. 2013, DellaGuistina et al. 2018). We extracted spectra from multiple image sets and measured the average spectra and standard deviation as error for normalized spectra (Fig. 2). As any I/F-based measurement needs to have a relative 5% error due to the radiometric calibration (Golish et al. 2020). And then it needs another 2% for the photometric correction error (Golish et al. 2021). Thus, in principle the absolute reflectance measurement has ~5.5% error in absolute value. We compared this value with standard deviation and apply the larger value as error of the reflectance factor (REFF) in Fig. 2. Some of exogenic boulders show linearly blue spectra in visible wavelength (v to x), which is not typical in anhydrous minerals. Despite this, those objects observed by the spectrometer are confirmed to have a 1-μm band (see section 2.3). Their spectra could be flattened by mixing or reacting with Bennu's carbonaceous composition, especially if the mixing timing was coincident with heating(see also Sec. 3 and 4.1).

## 2.3. OVIRS spectroscopy

The OSIRIS-REx Visible and InfraRed Spectrometer (OVIRS) (Reuter et al. 2018) is a hyperspectral, point spectrometer that measures the reflected and emitted energy of Bennu ranging wavelengths of 0.4 to 4.3 µm. During the Reconnaissance A (Recon A) phase of observations (Lauretta et al. 2017, 2021), OVIRS observed the OSIRIS-REx mission's four candidate sample sites, as well as some other areas of opportunity, with a footprint of ~4 m x 9 m (Simon et al. 2020). Some of the proposed exogenic materials were observed by the OVIRS instrument. Figure 3 shows the footprints of OVIRS during Recon A, which overlap with nine of the locations listed in Table 1. Four boulders (EX17, EX29, EX44, and EX72) show absorption bands around 1 and 2 µm in the OVIRS data. Because the OVIRS footprints are much larger than the exogenic objects, the reflectance spectra whose fields of view (FOV) overlapped them were divided by a global spectrum derived from all spectra acquired on the same date, to ensure similar illumination. This is particularly important in the case of small variations in the features. Figure 4 shows the OVIRS spectra divided by the global spectrum for these four boulders. EX44 and EX72 show very deep absorption bands at 1 and 2 µm. EX44 was detected as a bright boulder in DellaGiustina et al. (2021), and those absorption bands are consistent with calcium-poor pyroxenes (Simon et al. 2020). The mineralogy of EX44 and EX72 is discussed in Sec. 4.2.4. The other two, EX17 and EX29, show a weaker 1-µm absorption band and possible shallow absorption around 2 µm. The size of EX17 is ~0.3 m, whereas EX29 is several bright inclusions or clasts of sizes varying from 0.3 m to 0.5 m in a dark bedrock. The sizes of EX44 and EX72 are ~1.5 m and ~0.4 m, respectively. Although the total area of bright clasts in EX29 is >1 m$^2$ and in EX17 is ~1 m$^2$, which are much larger than areas occupied by EX72 inside of FOV, EX29 does not show a strong 2-µm band. This means that the shallowness of the absorptions observed in EX17 and EX29 is not due to their size. Thus, OVIRS spectra indicate compositional variation among the exogenic material on Bennu.

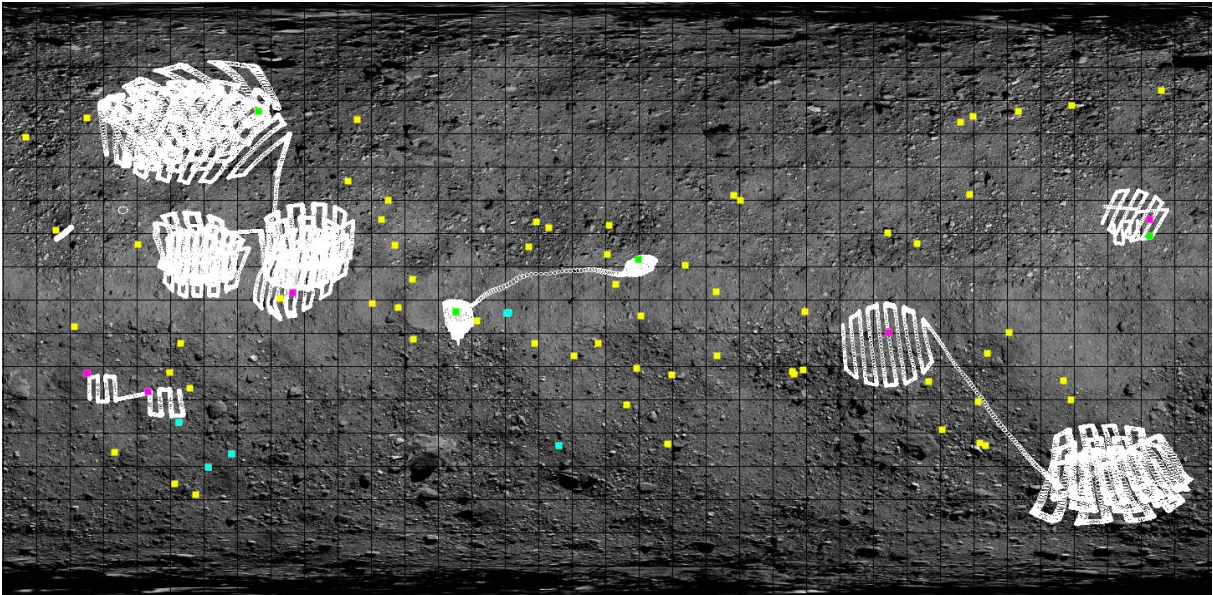

**Figure 3.** OVIRS footprints from Recon A are indicated by open white squares). Light blue squares indicate detection of 1- and 2-µm band absorptions reported in DellaGiustina et al. (2021) corresponding to the sites of exogenic material that we identified from MapCam data (Table 1). Green squares are the same except indicating detection during Recon A. Magenta squares indicate non-detection of these absorptions in OVIRS data overlying exogenic material, and yellow squares indicate sites of exogenic material that were not overlapped by the OVIRS footprints during Recon A.

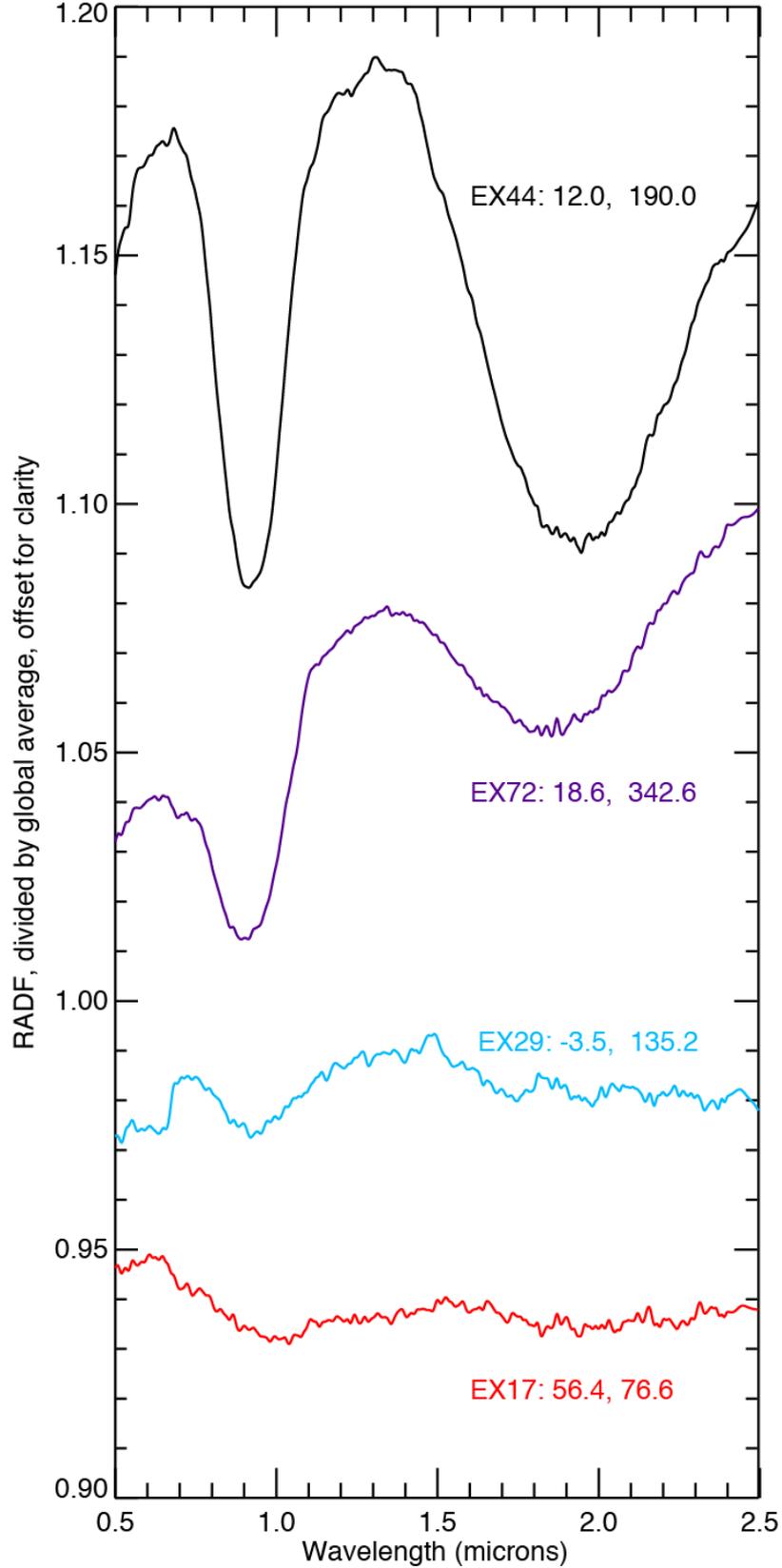

**Figure 4.** OVIRS spectra acquired for four instances of proposed exogenic materials. The spectra are generated by averaging together any overlapping OVIRS spots (areal resolution of ~40 m² per spot). Each spectrum is converted to radiance factor and divided by the average of all other OVIRS spectra acquired on that date. For EX44 and EX72 (top curves), the strong pyroxene signature is evident without this division,

but for EX29 and EX72 (bottom curves), the absorptions are weaker and only appear when divided by the global average. The spectrum of EX44 is the same spectrum shown in Fig. 1 of Simon et al. (2020).

## 3. Analyses
### 3.1. Comparison with other boulders on Bennu

The spectrophotometry of more than 1600 boulders were investigated in Fig. 2 of DellaGiustina et al. (2020). We conducted conversion from the normal condition to the standard condition, i.e., (i, e, $\alpha$)=(0°, 0°, 0°) to (30°, 0°, 30°), to compare with our results using the photometric model by Golish et al. (2021). They found four groups of boulders: dark, comparatively bright, Fe-bearing, and pyroxene-bearing (see Table 1 in DellaGiustina et al. (2020)). When plotting the b´ to x slope against the reflectance factor and the near-UV index (b´/v), DellaGiustina et al. (2020) found two major groups of boulders differentiated by their reflectance (dark versus bright) and a minor group of Fe-bearing phyllosilicate boulders, as well as pyroxene-bearing boulders as initially described in DellaGiustina et al. (2021). Median values of reflectance factor at standard geometry for dark boulder and comparatively bright boulder groups are 2.0% and 2.5%, and median values of $D_{xabs}$ are –1.6% and 1.0%, respectively. Although the "bright" boulders are distinct from the dark population, they are still relatively close to the average Bennu reflectance, and are a distinct group from the very bright objects that we propose to be exogenic. We plotted our color measurements of proposed exogenic objects with boulder data from DellaGiustina et al. (2020) for comparison (Fig. 5). The exogenic materials are significantly brighter, with reflectance factor at standard geometry ranging from 2.8% to 8.8% (Fig. 5a), than most of the boulders on the surface of Bennu and have considerably lower values for the near-UV index, i.e., deeper absorption towards the near-UV (Fig. 5c) and 1 µm (Fig. 5b). This significant difference in brightness of the proposed exogenic objects from the main populations of boulders on Bennu supports their origin being from a distinct parent body or bodies.

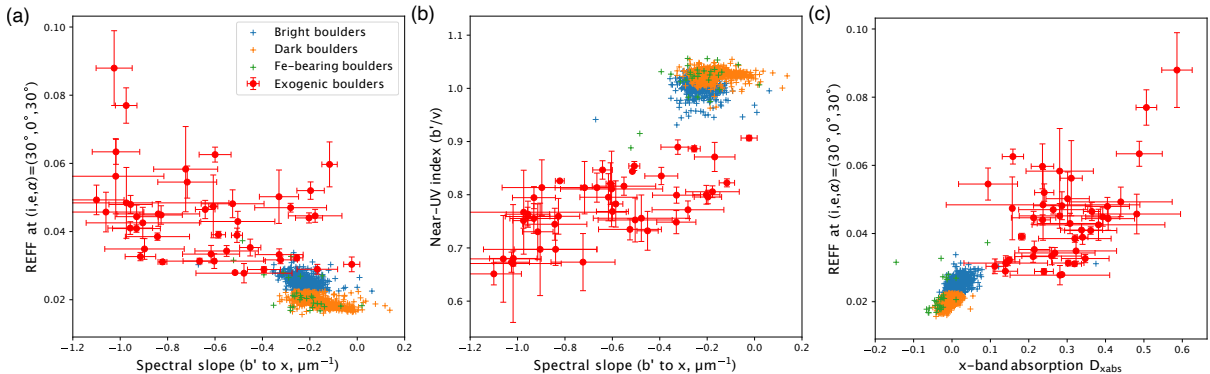

**Figure 5.** Color distribution of proposed exogenic objects from this study (red points) and other boulders (blue, green, and orange crosses) mapped by DellaGiustina et al. (2020). (a) Spectral slope (b´ to x) versus reflectance factor; (b) near-UV index; (c) x-band absorption and reflectance factor.

### 3.2. Principal component analysis

The proposed exogenic materials on Bennu show diversity in spectra and morphology. We conduct principal component analysis (PCA) to extract the maximum variation among their spectrophotometry. PCA has been often used to classify main belt asteroids (e.g., Tholen, 1984; Bus and Binzel, 2002). The purpose of PCA is to search for a sequence of principal component vectors to minimize the average squared distance from the dataset points to the vector line, while the vectors need to be orthogonal to each other. We used the spectra of the 46 instances of exogenic material for which these data are available as input (Fig. 6).

The first principal component (PC1) is similar to the shape of the mean of 46-spectra dataset, suggesting that the largest source variation among the 46 spectra is albedo. The second principal component (PC2) represents variation in the depth of x-band and b´-band absorptions, i.e., a higher PC2 score suggests less absorption at x and b´ bands.

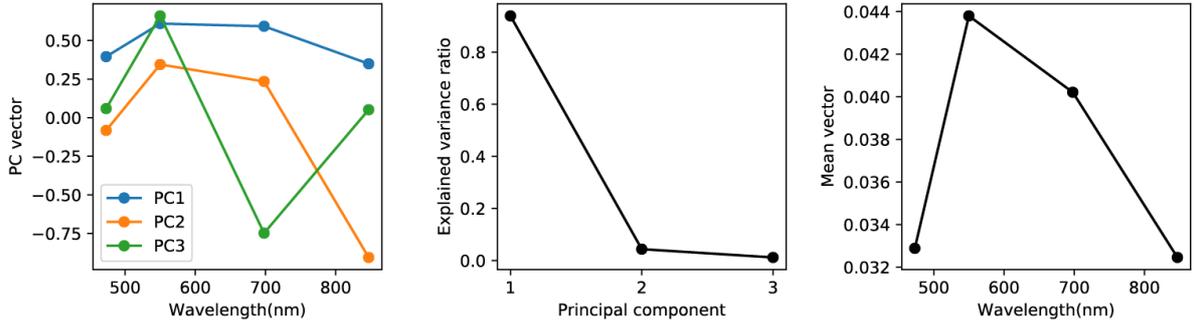

**Figure 6.** PCA of 46 spectra of exogenic materials. *Left*: Principal component vectors. *Middle*: Explained variance ratio is the same as the eigen-values, which indicate how much each vector can explain the whole variation in the dataset, showing the importance of each of the three components. *Right*: mean spectra of the dataset.

### 3.3. Morphologic expression of exogenic materials

Using PolyCam images, we classified the morphology of the proposed exogenic materials into four categories: homogenous, breccia-like, inclusion-like, and others (including boulders with many inclusions and boulders that are partially bright) (Table 2). Figure 7 shows examples of these morphological groups. Homogeneous rocks have almost the same brightness for the whole rock, while breccia-like rocks seem to be mixtures of similar amounts of bright and dark compositions. Inclusion-like features are small, bright clasts embedded in dark host rocks with cauliflower-like, rugged textures. The inclusions seem to be exposed by the erosion of the host rocks. Other morphologies include, for example, EX29, which has a large number of inclusions. EX71 seems to have a exposed, brighter facet on the surface (we call this "partially bright"). (Images of all proposed exogenic materials are shown in Fig. A1.)

We plotted the distribution of the morphologic groups in Fig. 8a. There is no strong correlation between locations and morphologies, suggesting that all morphological groups are well mixed with common materials on Bennu. To integrate the spectral and morphological information, we plotted PC1-PC2 with information from morphological classification in Fig. 9b. The brightest materials (high PC1 score) can be divided into two groups: homogeneous rocks with deep x-band absorptions (EX11, EX44, and EX27), and breccia-like rocks with shallow x-band absorptions (EX36 and EX51). In the PC1-PC2 space, two dominant trends can be seen: one from Bennu's average to EX11, and another from Bennu's average to EX36 (marked as Trend I and Trend II in Fig. 8b, respectively). Thus, EX11 and EX36 could be considered as endmembers and the rest of the points as mixtures between the endmember materials and Bennu's average. Both EX11 and EX36 are much brighter than the average, while EX11 shows more x- and b´- band absorption. DellaGiustina et al. (2021) showed that EX11 is spectrally similar to HED meteorites. However, they did not propose spectral similarity of EX36 to HEDs due to its low signal-to-noise ratio, meaning that EX36 could be a different kind of anhydrous-silicate-rich material. The PC1 (Fig. 8b), spectroscopic (Fig. 4), and morphological (Fig. 7) differences between EX11 and EX36 suggest that they are compositionally different. Moreover, along Trend I, EX44 and EX72 were observed by OVIRS and show deep 1- and 2-μm band absorptions (Fig. 4), which is consistent with basaltic material. On the other hand, along Trend II, only EX29 was observed by OVIRS and shows shallower 1- and 2-μm bands (Fig. 4), which is consistent with olivine-rich material.

Along with Trend I, we can see the transition of morphology from homogeneous bright rocks to darker inclusion-like or breccia-like objects. This also supports the hypothesis of mixing between Bennu's dominant composition and pyroxene-rich exogenic materials.

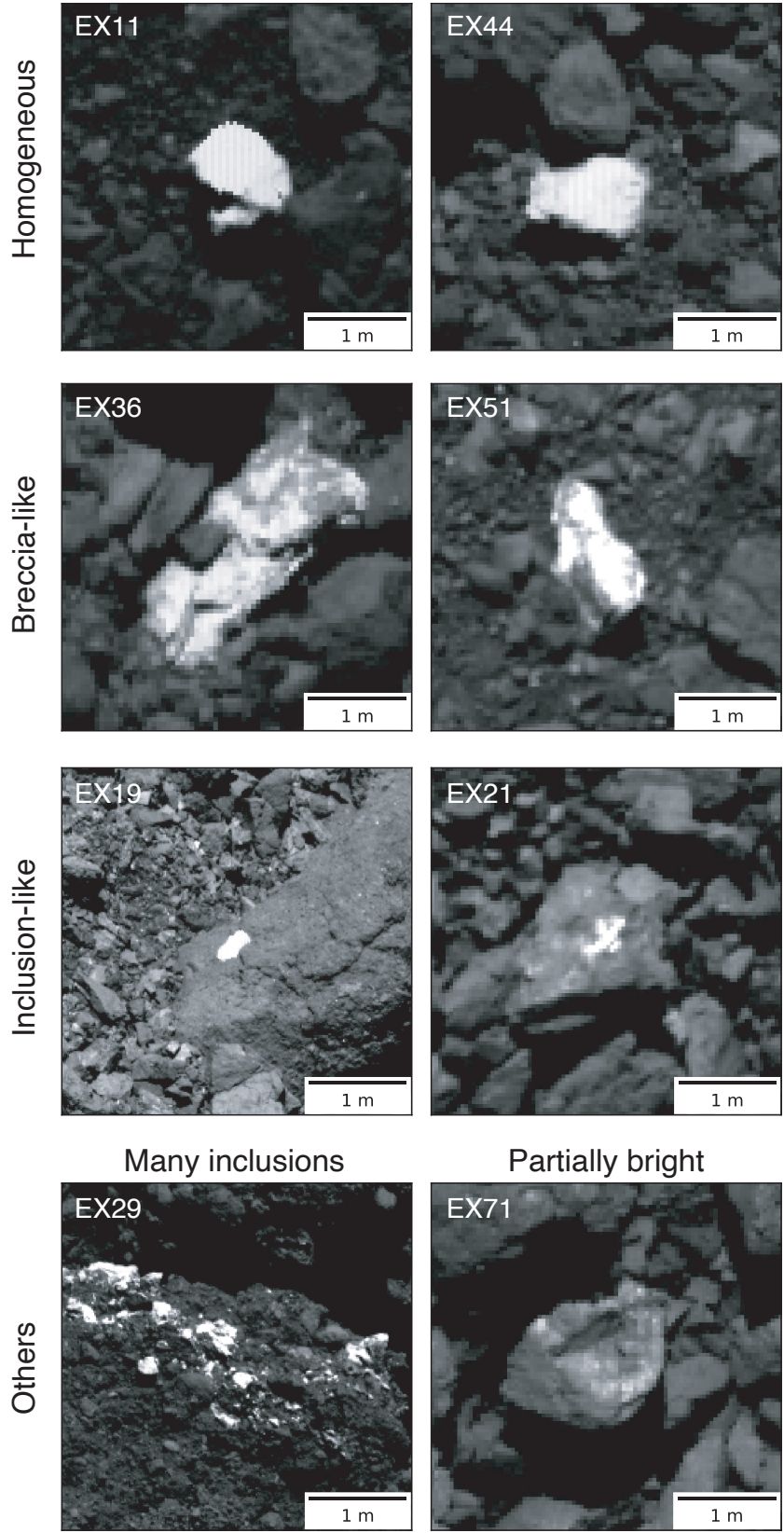

**Figure 7.** Examples of each morphological classification of the exogenic materials using PolyCam images. Image IDs: 20190405T210531S863 (3.8 cm/pixel) for EX11, 20190321T184411S010 (4.8 cm/pixel) for EX44, 20190321T190056S516 (4.8 cm/pixel) for EX36, 20190412T183616S181 (3.8 cm/pixel) for EX51, 20191019T210433S861 (1.5 cm/pixel) for EX19, 20190412T195755S393 (3.8 cm/pixel) for EX21, 20191026T201824S739 (1.6 cm/pixel) for EX29, and 20190412T172747S061 (3.8 cm/pixel) for EX71.

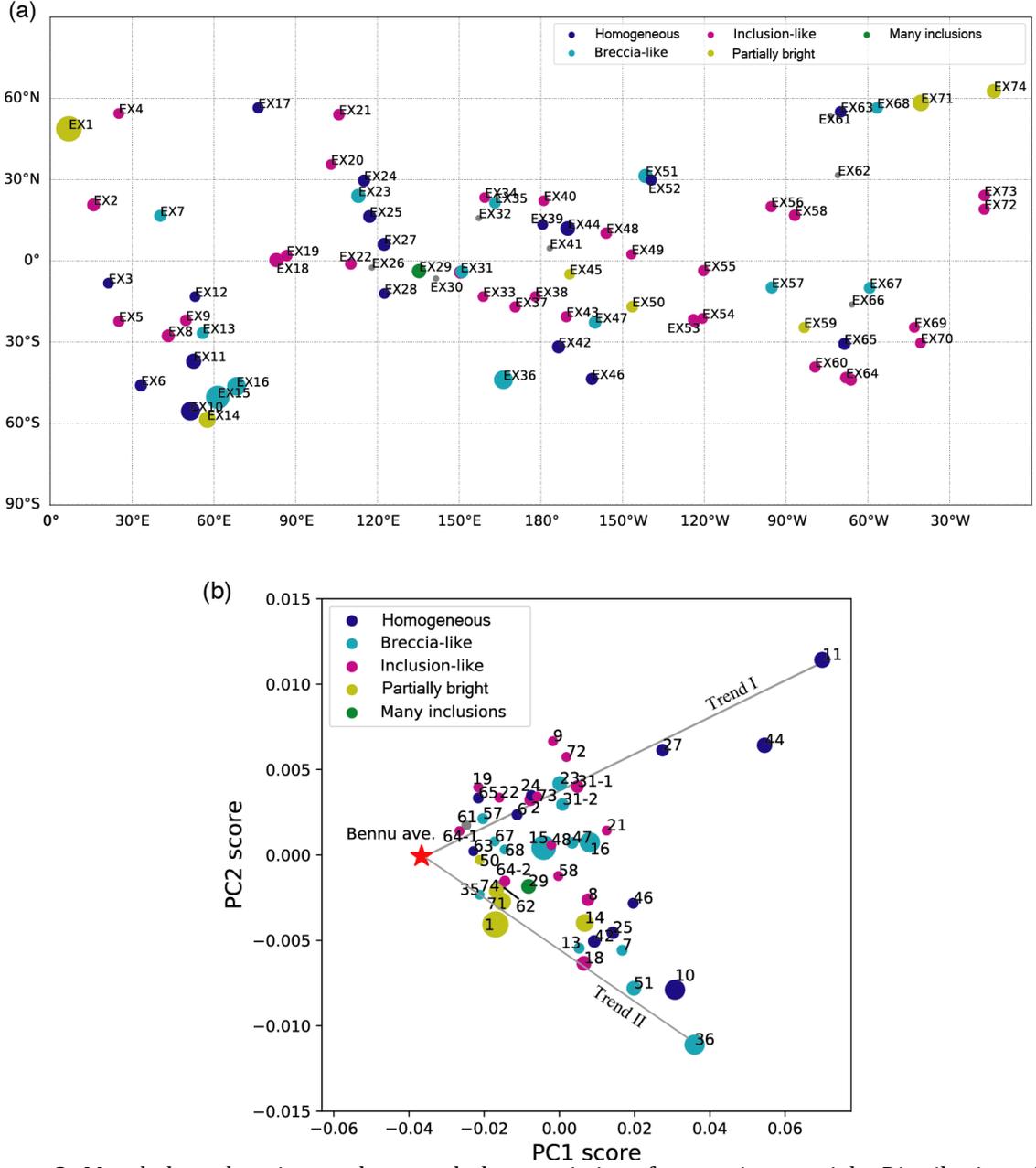

**Figure 8.** Morphology, location, and spectral characteristics of exogenic materials. Distribution of 46 spectra in PC1-PC2 space. The red star indicates the average spectrum of Bennu. The numbers correspond to the identification numbers (EX#) of the exogenic materials shown in Table 1 and Fig. A1. (a) Location and morphology. (b) PC scores and morphology. The size of the marker is set according to the relative size of the objects.

**Table 2.** Morphology and spectral characteristics of exogenic materials. (REFF is a reflectance factor.)

| ID  | REFF (%) at (30°,0°,30°) | Morphology      | OVIRS          | ID    | REFF (%) at (30°,0°,30°) | Morphology      | OVIRS              |
|-----|--------------------------|-----------------|----------------|-------|--------------------------|-----------------|--------------------|
| EX1 | 3.0±0.2                  | Partially bright|                | EX39  |                          | Homogeneous     |                    |
| EX2 | 4.1±0.2                  | Inclusion-like  |                | EX40  |                          | Inclusion-like  |                    |
| EX3 |                          | Homogeneous     |                | EX41  |                          |                 |                    |
| EX4 |                          | Inclusion-like  |                | EX42  | 4.7±0.3                  | Homogeneous     |                    |
| EX5 |                          | Inclusion-like  | No 2-μm band   | EX43  |                          | Inclusion-like  |                    |
| EX6 | 3.9±0.2                  | Homogeneous     |                | EX44* | 7.7±0.5                  | Homogeneous     | Strong 2-μm band   |
| EX7 | 5.5±0.5                  | Breccia-like    |                | EX45  |                          | Not very bright |                    |
| EX8 | 4.8±0.4 | Inclusion-like | No 2-µm band | EX46 | 5.8±1.2 | Homogeneous | |
| EX9 | 4.6±0.6 | Inclusion-like | | EX47 | 4.8±1.0 | Breccia-like | |
| EX10 | 6.3±0.3 | Homogeneous | | EX48 | 4.5±0.4 | Inclusion-like | |
| EX11* | 8.8±1.1 | Homogeneous | | EX49 | | Inclusion-like | |
| EX12 | | Homogeneous | | EX50 | 3.1±0.2 | Not very bright | |
| EX13 | 4.4±0.2 | Breccia-like | | EX51 | 5.2±0.3 | Breccia-like | |
| EX14 | 4.7±0.9 | Partially bright | | EX52 | | Homogeneous | |
| EX15* | 3.9±0.2 | Breccia-like | 2-µm band† | EX53-1 | | Inclusion-like | |
| EX16* | 4.6±0.3 | Breccia-like | 2-µm band† | EX53-2 | | Inclusion-like | |
| EX17 | | Homogeneous | Possible 2-µm band | EX54 | | Inclusion-like | |
| EX18 | 4.5±0.2 | Inclusion-like | | EX55 | | Inclusion-like | |
| EX19 | 3.3±0.2 | Inclusion-like | No 2-µm band | EX56 | | Inclusion-like | |
| EX20 | | Inclusion-like | | EX57 | 3.1±0.2 | Breccia-like | No 2-µm band |
| EX21 | 5.6±1.1 | Inclusion-like | | EX58 | 4.3±0.3 | Inclusion-like | |
| EX22 | 3.5±0.3 | Inclusion-like | | EX59 | | Partially bright | |
| EX23 | 4.4±0.3 | Breccia-like (Inclusion-like) | | EX60 | | Inclusion-like | |
| EX24 | 4.1±0.2 | Homogeneous | | EX61 | | | |
| EX25 | 5.0±0.8 | Homogeneous | | EX62 | 3.3±0.2 | | |
| EX26 | | | | EX63 | 2.9±0.2 | Homogeneous | |
| EX27 | 6.3±0.4 | Homogeneous | | EX64-1 | 2.8±0.3 | Inclusion-like | |
| EX28 | | Homogeneous | | EX64-2 | 3.5±0.2 | Inclusion-like | |
| EX29 | 3.9±0.2 | Many inclusions | Possible 2-µm band | EX65 | 3.1±0.2 | Homogeneous | |
| EX30 | | | | EX66 | | Partially bright | |
| EX31-1* | 4.8±0.3 | Inclusion-like | 2-µm band† | EX67 | 3.3±0.3 | Breccia-like | |
| EX31-2* | 4.5±0.3 | Breccia-like (Inclusion-like) | 2-µm band† | EX68 | 3.4±0.2 | Breccia-like | |
| EX32 | | Partially bright | | EX69 | | Inclusion-like | |
| EX33 | | Inclusion-like | | EX70 | | Inclusion-like | |
| EX34 | | Inclusion-like | | EX71 | 3.2±0.2 | Partially bright | |
| EX35 | 2.9±0.2 | Breccia-like | | EX72 | 4.9±0.4 | Inclusion-like | Strong 2-µm band |
| EX36* | 6.0±0.7 | Breccia-like | 2-µm band† | EX73 | 4.3±0.5 | Inclusion-like | No 2-µm band |
| EX37 | | Inclusion-like | | EX74 | 3.2±0.2 | Partially bright | |
| EX38 | | Inclusion-like | | | | | |

†Observed by OVIRS with lower spatial resolution (DellaGiustina et al. 2021).

### 3.3. Comparison with laboratory data

To constrain the possible compositions represented by the spectrophotometric data of the proposed exogenic materials, we compared the 46 spectra with laboratory data of meteorites and minerals. We used spectra from the RELAB[2] database in its December 29, 2019 version as a reference set.

We selected the spectra of meteorites and minerals that include the wavelength interval covered by the b´, v, w, and x filters of the MapCam instrument. Only those laboratory data obtained at a phase angle of 30° are used. Correspondingly, the average reflectance factors at ($i$,$e$, $\alpha$ ) = (30°,0°,30°) of Bennu's exogenic objects are used for computation. The resulting set of data used for comparison includes 1986 spectra of meteorites and 8820 spectra of other terrestrial materials.

The preliminary step is to transform the spectral curves obtained in the laboratory to the MapCam filter system. This is done by multiplying each spectrum by the four filter functions (Rizk et al. 2018), then integrating over each bandpass to obtain the total reflection in the corresponding band-interval.

The first approach is to perform a direct comparison between the MapCam data and those corresponding to meteorites. This approach may provide a first rough guess about the boulders' compositions, which can drive the further analysis. The matching algorithm takes into account the spectral shapes. The wide spectral features of asteroids can be identified even using the broad-band filters. However, before applying the curve-matching algorithm, a pre-selection is made. For each exogenous candidate, we chose the laboratory data that have the same absolute reflectance value in the v-filter, within ±3σ. Although the absolute reflectance value is affected by various errors (including unaccounted errors due to the calibration process), we prefer to use this information by considering its uncertainty in order to obtain a meaningful comparison. Then, all spectrophotometric data are normalized to the v-filter and the least squared differences are computed between RELAB spectra and the MapCam spectra. The best matches are found for each bright boulder.

[2] http://www.planetary.brown.edu/relab/

A reasonable spectral match was found for EX24, which does not have a deep band in the x filter (i. e., large w/x ratio). Thus, it resembles some CI and heated CM heated samples (Fig. 9). Also, the bright boulders showing a peak in the v filter have a similar spectral shape to several spectra of CM2 carbonaceous chondrites (an example is shown in Fig. B1). This fitting is precluded by the deep features in the b´ and x filters and cannot be accepted as a solution.

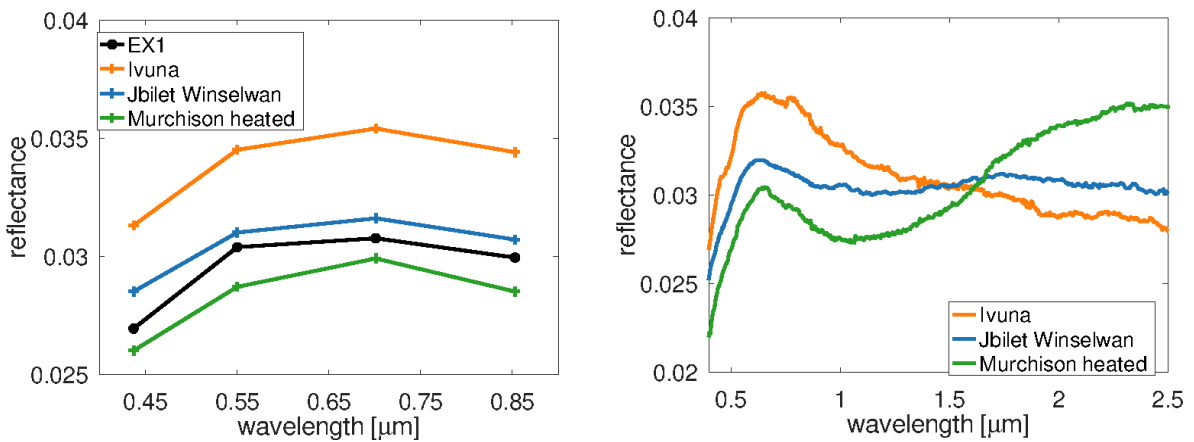

**Figure 9**. *Left*: MapCam spectrum of bright boulder EX1 (black) *versus* spectrophotometric data of CI Ivuna (SampleID: MB-TXH-060, FileID: C4MB60, in orange), CM2 Jbilet Winselwan (SampleID: MT-DAK-312-B, FileID: C1MT312B, in blue), and Murchison heated at 500° (SampleID: MB-TXH-064-HI, FileID: CIMB64, in green). *Right*: the Relab visible-to-near-infrared spectra of the three meteorites.

For the rest of the boulders the results of direct comparison give unsatisfactory matches (i. e. large differences between the spectral shapes). In some cases several best-matches were found, but these are of uncommon samples. These include spectra of the Esquel pallasite meteorite (fileID: C6MB43, SampleID: MB-TXH-043); an example is plotted in Fig. B2. According to the information provided by the RELAB database, this meteorite sample was processed in the laboratory as follows: "side 1 polished with sandpaper #60, side 2 polished like a mirror", which is not probable in natural environments. We note that several spectra are available for this meteorite and there are substantial differences between them (some of the spectral curves do not show the 1-µm band, and the absolute reflectance value varies between 0.005 and 0.30); thus, this cannot be considered a reliable match. These results are a hint for the peculiarity of the bright-boulder spectrophotometric curves.

We must take into account that Bennu's exogenic materials are likely a mixture of olivine and pyroxene with matrix material, including carbon content that lowers albedo. In fact, the prominent spectral features expressed by Bennu's exogenic materials in the b´ and x filters are not matched by the RELAB data when the absolute reflectance value of the b´ filter is taken into account. The deep absorption band inferred from the (x-w) color suggests a pyroxene/olivine mineralogy. Such compositions have in most cases a much higher reflectance than that found for the exogenic materials on Bennu.

We also note the differences in the size of the samples for which the spectrophotometric curves are compared. The binned MapCam data correspond to a resolution on the order of 50 cm, whereas the RELAB spectra are of samples with diameters on the order of a millimeter. Therefore, spectral mixtures of various laboratory samples may provide an analogue for the data observed for Bennu's bright boulders. There are two main approaches for this method (Reddy et al. 2015). First, the approximation of an areal mixture ("checkerboard") consists of a linear combination of reflectance spectra of various constituents to represent the total reflectance. The second model corresponds to intimate homogeneous mixing ("salt and pepper") and involves computing an average single-scattering albedo of the mineral constituents as input to the reflectivity calculation (e.g. Shkuratov et al., 1999).

The PolyCam images suggest the use of an areal mixture model is appropriate. The bright patches are well defined with clear edges on the dark Bennu terrain. Also, a simple linear mixture approach is effective against the errors of the spectrophotometric data. Thus, we artificially generated new spectra following the formula:

$$Spec_{artif} = coef1 * Spec1 + coef2 * Spec2$$

where coef1 + coef2 = 1. $Spec1$ is the average reflectance factor of Bennu (b' = 0.01986, v = 0.019794, w = 0.019477, x = 0.019226) while Spec2 is one the 1986 meteorite spectra. Thus, we were able to generate a database of 196,614 artificial spectra by varying the coef1 between 0.01 and 0.99 with a step size of 0.01. With this database of simulated spectra, we performed the same curve-matching algorithm as before.

These areal mixtures can explain the majority of the spectrophotometric data found for the proposed exogenic materials on Bennu. In most cases, it involves a mixture of 85–95 % of the average spectrum of Bennu with 15–5 % of the spectrum of HEDs. However, even when using these linear mixtures, the exogenic objects with a spectral peak at 0.55 µm and a steep drop in the w and x filters remain difficult to explain.

One of the combinations matching this spectral shape is obtained by mixing the spectrum of the eucrite Macibini Cl.3 melted clast (Buchanan et al. 2000) with the average spectrum of Bennu (Fig. 10). This sample is from clast 3 of Macibini Cl.3, which is an impact melt breccia that is composed of rock and mineral fragments in a devitrified groundmass. Its diameter is about 2-cm and it is similar to other polymict eucrites. Buchanan et al (2000) suggested that it may represent an impact melting of the same regolith material represented in the matrix of the meteorite. Given that boulders of HED-like composition are known to be present on Bennu, the detection of impact melts with a composition expected from HED impacts could constrain impact models (e.g., Daly et al. 2016).

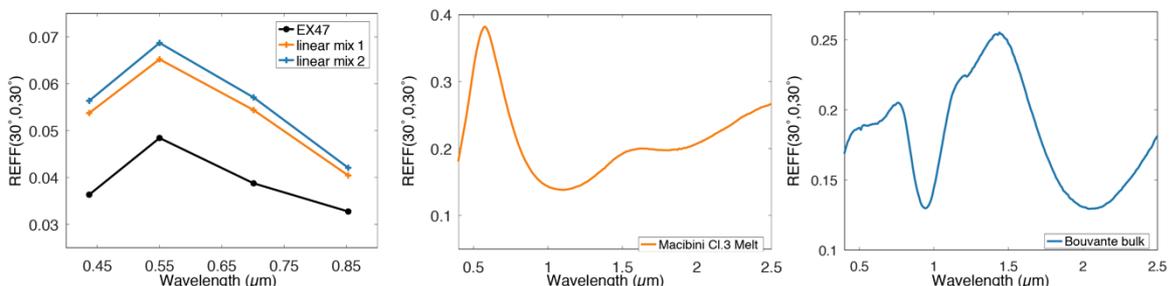

**Figure 10.** *Left*: MapCam spectrum of bright boulder EX47 (black) *versus* mixtures between average Bennu and 13% of eucrite Macibini CI.3 melted clast (SampleID: TB-RPB-030, FileID: C1TB30, in orange. An alternative solution (in blue) is obtained using also 14% of eucrite Bouvante bulk (SampleID: TB-RPB-029, FileID: C1TB29). Visible-to-near-infrared spectra of the two meteorite samples are shown in the central and right panels.

The solutions presented above are most representative. But the curve-matching comparisons show a variety of solutions which are valid within the error bars. A generalized approach is to consider the comparison in the principal components space defined in Section 3.2. We therefore projected spectra generated as linear mixtures between the average Bennu and the laboratory spectral curves of meteorites in this space (Fig. 11). The aim of this comparison is to look for similarity with the most common types of meteorites. Thus, we took into account 280 spectra of HED, 440 spectra of carbonaceous chondrite, and 414 spectra of ordinary chondrites. The regions occupied by these meteorites (defined as ±2σ from the average values) are shown in Fig. 11. We notice that there is a slight overlap between the three regions.

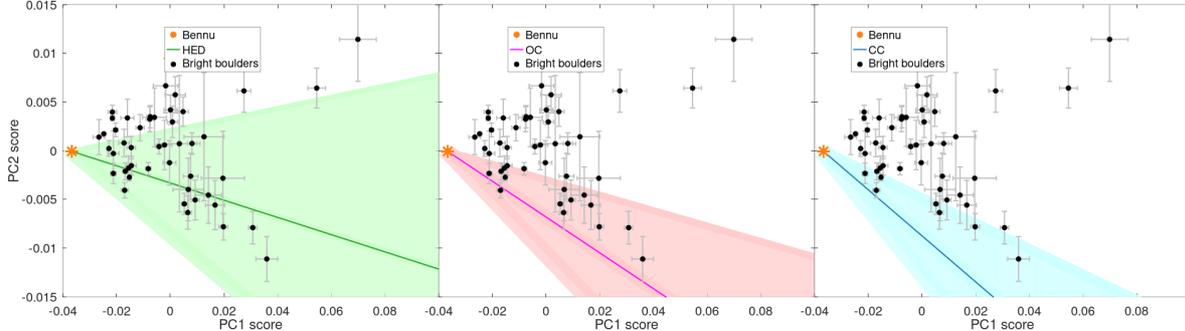

**Figure 11**. Comparison between the principal components of exogenic objects (black points with the error-bars shown in gray) and the regions occupied by the data corresponding to linear mixtures between average Bennu's spectrum and the laboratory spectral curves of meteorites for HEDs (green region), ordinary chondrites (red region), and carbonaceous chondrites (blue region). Lines indicate the median mixture spectra and hatches indicate the region of 2 sigma variation.

Figure 11 shows the variety of compositions that can explain the spectrophotometric behaviour of different exogenic rocks on Bennu. In the PC1 versus PC2 space, the boulders occupy the same region as the linear mixtures between Bennu's spectrum and meteorites. Some of the boulders, especially the ones on Trend I, can be explained by linear mixtures of the Bennu spectrum with those of HED meteorites, while for others among Trend II, we cannot exclude the presence of olivine, pyroxene, and ordinary-chondrite–like compositions. Le Corre et al. (2021) also suggest the possibility of mixture between ordinary chondrites and Bennu's typical material for a broader population (~170) of larger boulders (PYR1b in their classification). Moreover, some exogenic objects in Trend II can also be explained by mixtures with carbonaceous chondrite compositions. In that case, we cannot exclude an endogenous origin of those materials. The fact that some of the points are just outside the regions defined in Fig. 11 for the various mixtures of meteorites highlights their peculiar composition, especially the strong 1-μm band absorption. However, it is also true that there are a few HED examples, outside of 2-sigma, to account for the mixing Trend I.

A probability for the matching with meteorite types can be computed by considering the regions defined in Fig. 11 and the error bars from the exogenic materials' spectrophotometric data. This was computed using a Monte Carlo approach. We generated 1E6 spectrophotometric clones for each exogenic object. This was done using normally distributed random points, where the averages are the corresponding PC1 and PC2 values, and the standard deviations are the errors. Then, for each clone we found the region where it belongs. The probability for an exogenic object to be classified as HED-, OC-, or CC-like is the numbers of clones found in the corresponding region divided by the total number of clones (1E6). The results are shown in Table 3, and they outline the similarity with HED meteorites for most of our proposed exogenic materials. The sum of these probabilities is not 1 because of the overlapping between the regions. Also, the points falling in regions that overlap may have the probability 1 for different groups (they cannot be distinguished with our data). The probabilities shown in Table 3 provide a quantitative measurement for the resemblance with meteorite groups, and they also quantify the effect of errors in this comparison.

**Table 3.** The probability of each boulder having a composition similar to an real mixture between Bennu's average and HED, OC, or CC meteorites.

| BoulderID | prob$_{HED}$ | prob$_{OC}$ | prob$_{CC}$ | BoulderID | prob$_{HED}$ | prob$_{OC}$ | prob$_{CC}$ |
|---|---|---|---|---|---|---|---|
| EX1 | 0.77 | 1 | 0.96 | EX35 | 0.9 | 0.84 | 0.59 |
| EX2 | 1 | 0 | 0 | EX36 | 0.99 | 0.99 | 0.74 |
| EX6 | 1 | 0 | 0 | EX42 | 1 | 0.74 | 0.31 |
| EX7 | 1 | 0.69 | 0.28 | EX44 | 1 | 0 | 0 |
| EX8 | 1 | 0.34 | 0.09 | EX46 | 0.98 | 0.36 | 0.18 |
| EX9 | 1 | 0 | 0 | EX47 | 0.92 | 0.29 | 0.2 |
| EX10 | 1 | 0.93 | 0.28 | EX48 | 1 | 0.09 | 0.02 |
| EX11 | 1 | 0 | 0 | EX50 | 0.93 | 0.33 | 0.22 |
| EX13 | 1 | 0.97 | 0.46 | EX51 | 1 | 0.99 | 0.6 |
| EX14 | 0.96 | 0.56 | 0.3 | EX57 | 1 | 0 | 0 |
| EX15 | 1 | 0 | 0 | EX58 | 1 | 0.23 | 0.06 |
| EX16 | 1 | 0 | 0 | EX61 | 1 | 0 | 0 |
| EX18 | 0.99 | 0.95 | 0.64 | EX62 | 0.99 | 0.51 | 0.2 |
| EX19 | 1 | 0 | 0 | EX63 | 1 | 0.04 | 0 |
| EX21 | 0.97 | 0.21 | 0.12 | EX64-1 | 0.98 | 0.1 | 0.06 |
| EX22 | 1 | 0 | 0 | EX64-2 | 1 | 0.38 | 0.07 |
| EX23 | 1 | 0 | 0 | EX65 | 1 | 0 | 0 |
| EX24 | 1 | 0 | 0 | EX67 | 1 | 0.05 | 0.01 |
| EX25 | 0.99 | 0.56 | 0.23 | EX68 | 1 | 0.03 | 0 |
| EX27 | 1 | 0 | 0 | EX71 | 1 | 0.99 | 0.34 |
| EX29 | 1 | 0.22 | 0 | EX72 | 1 | 0 | 0 |
| EX31-1 | 1 | 0 | 0 | EX73 | 1 | 0.02 | 0.01 |
| EX31-2 | 1 | 0 | 0 | EX74 | 1 | 0.7 | 0.26 |

*4.2.4 Mineralogical analysis*

The spectral band parameters of olivine-pyroxene compositions can be correlated with the mafic mineral abundances (Cloutis et al. 1986; Gaffey et al. 1993; Dunn et al. 2013). An in-depth review of this technique is provided by Reddy et al. (2015), who provided a flow-chart for mineralogical characterization of A-, S-, and V-type asteroids. We applied this algorithm using as input the OVIRS spectra of EX44 and EX72, which have sharp absorption bands at 1 and 2 μm. As shown in the previous section, these spectra correspond to a mixture between the common carbonaceous material of Bennu and a small fraction of various pyroxene-olivine compositions. As a consequence, the brightness of these areas and the depth of the spectral bands are greatly reduced. As a rough approximation, we assume that the presence of carbonaceous material does not modify the band shapes meaningfully (such that the empirical models from the laboratory are still valid).

Another case of mixing HED-like material with CC compositions is reported by De Sanctis et al. (2015). They used the data provided by Dawn mission for the Marcia region from the asteroid (4) Vesta. They report that the band parameters of HED-like compositions identified in this region appear modified when they were mixed with "exogenic" CC-like material.

**Table 3.** The band parameters computed for the OVIRS spectra of EX44 and EX72. BImin and BIImin represent the first band (BI) and second band (BII) minima; the BIC and BIIC are the band centers. BAR represents the band area ratio.

| #ID  | BI$_{min}$ | BI$_{min}$_err | BII$_{min}$ | BII$_{min}$_err | BIC   | BIC_err | BIIC  | BIIC_err | BAR   | BAR_err |
|------|------------|----------------|-------------|-----------------|-------|---------|-------|----------|-------|---------|
| EX44 | 0.950      | 0.018          | 1.895       | 0.095           | 0.958 | 0.018   | 1.897 | 0.095    | 1.765 | 0.301   |
| EX72 | 0.947      | 0.015          | 1.905       | 0.095           | 0.946 | 0.015   | 1.875 | 0.095    | 2.106 | 0.132   |

First, we computed the wavelengths of the reflectance minima around 1 µm and 2 µm (BImin and BIImin), the band centers (BIC and BIIC, after continuum removal), and the band area ratio (BAR), which is the ratio of the areas of the second and the first absorption band. If there is no overall continuous slope in the spectrum, the band center and the band minimum are coincident. Otherwise, a continuum slope in the spectral region of the absorption feature will shift the band center by an amount related to the slope of the continuum and the shape of the absorption feature. The results are shown in Table 3.

The computation of band minima is affected by spectral artifacts on the recorded spectrum. The computation was made by performing various polynomial fittings of the spectral curves around the two minima. Different polynomial degrees and different spectral intervals were considered to perform the fit. The results is the median value of all these test, and the error is their standard deviation.

To illustrate these results, a zoomed-in view of these bands and the estimated minima are outlined in Fig. 12. An additional error is introduced by the poor quality of the two spectra around 0.7 µm. Th precludes the accurate evaluation of the continuum (computed as a linear slope between the maximum at 0.7 µm and the near-infrared one), and introduces an error of ±0.015 to 0.018 µm for the BIC and of about 0.095 µm for the BIIC.

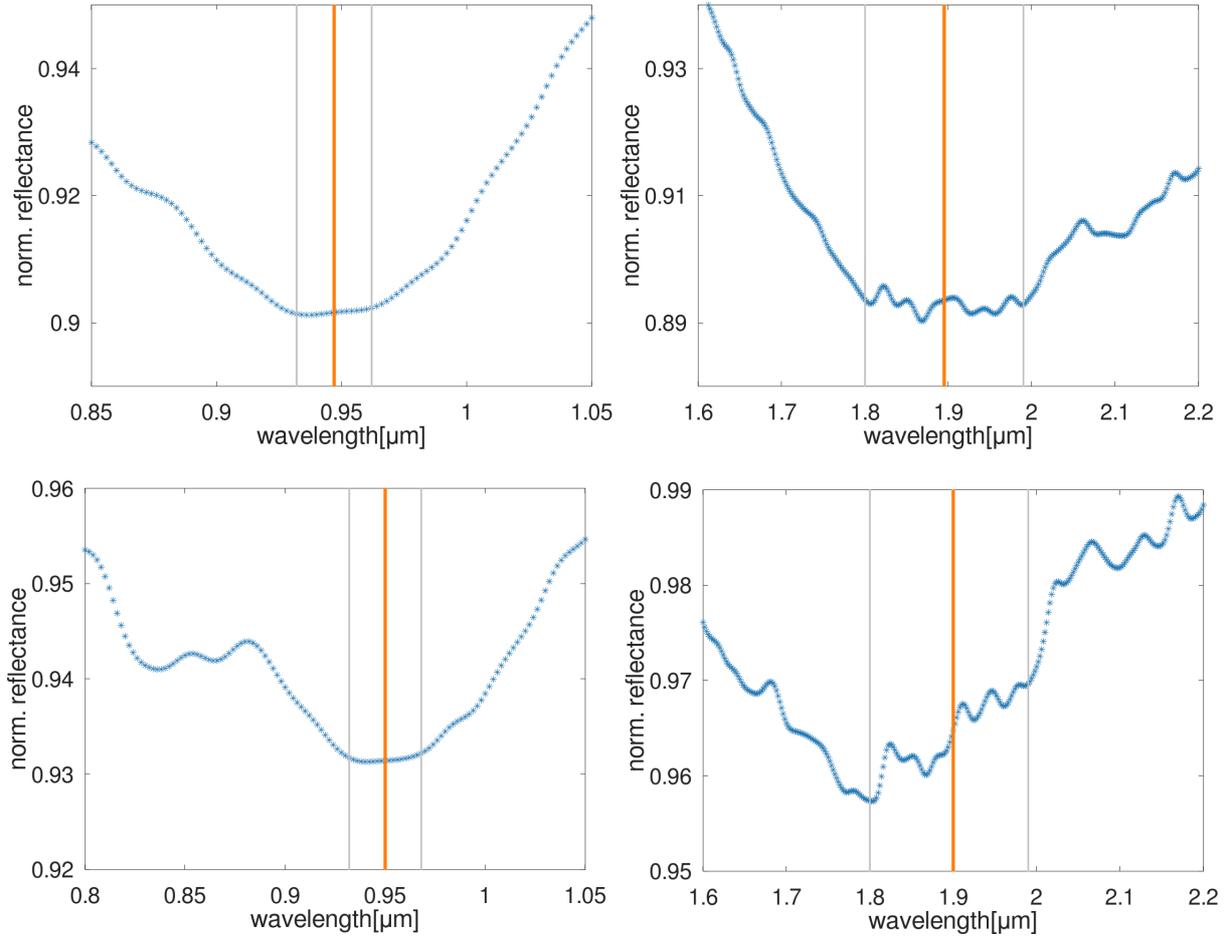

**Figure 12.** Zoom-in of the position of the two band minima. The two top panels present the spectrum of EX72, and those of the bottom show the spectral curves of EX44. The red vertical line shows the determined BImin, and the gray one highlights the errors.

These methods were mainly developed to study main belt asteroids, so have to apply a temperature correction to account for the higher surface temperature of a near-Earth asteroid like Bennu. By taking into account the equations shown by Reddy et al. (2015) and references therein, the variation of band centers are of the order ΔBIC = ±0.001 μm, ΔBIIC = ±0.012 μm for a temperature between 200 and 400 K.

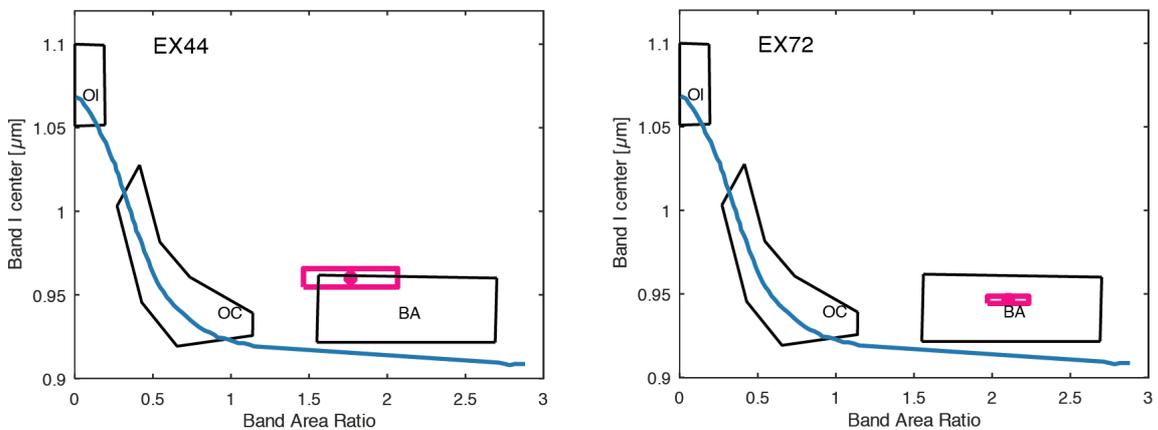

**Figure 13.** BAR versus BIC for the two analyzed boulders (red). Compositional regions defined by Dunn et al. (2013), following the results from Gaffey et al. (1993) and Burbine et al. (2001), are also displayed,

where O corresponds to olivine, OC to ordinary chondrites, and BA to basaltic achondrites. The blue solid line indicates the location of the olivine-orthopyroxene mixing line.

The plot shown in Fig. 13 indicates a composition similar to basaltic achondrites for the two boulders. The position of the BIC and BIIC centers, compared to those of synthetic pyroxenes (Fig. 14), place them in the low-calcium pyroxene region, which is consistent with the findings of DellaGiustina et al. (2021). The large error bars preclude a confident matching with one of the HED types. Those two features are both components of Trend I, suggesting that this trend is probably caused by the mixture between basaltic HED and Bennu's mainly carbonaceous composition.

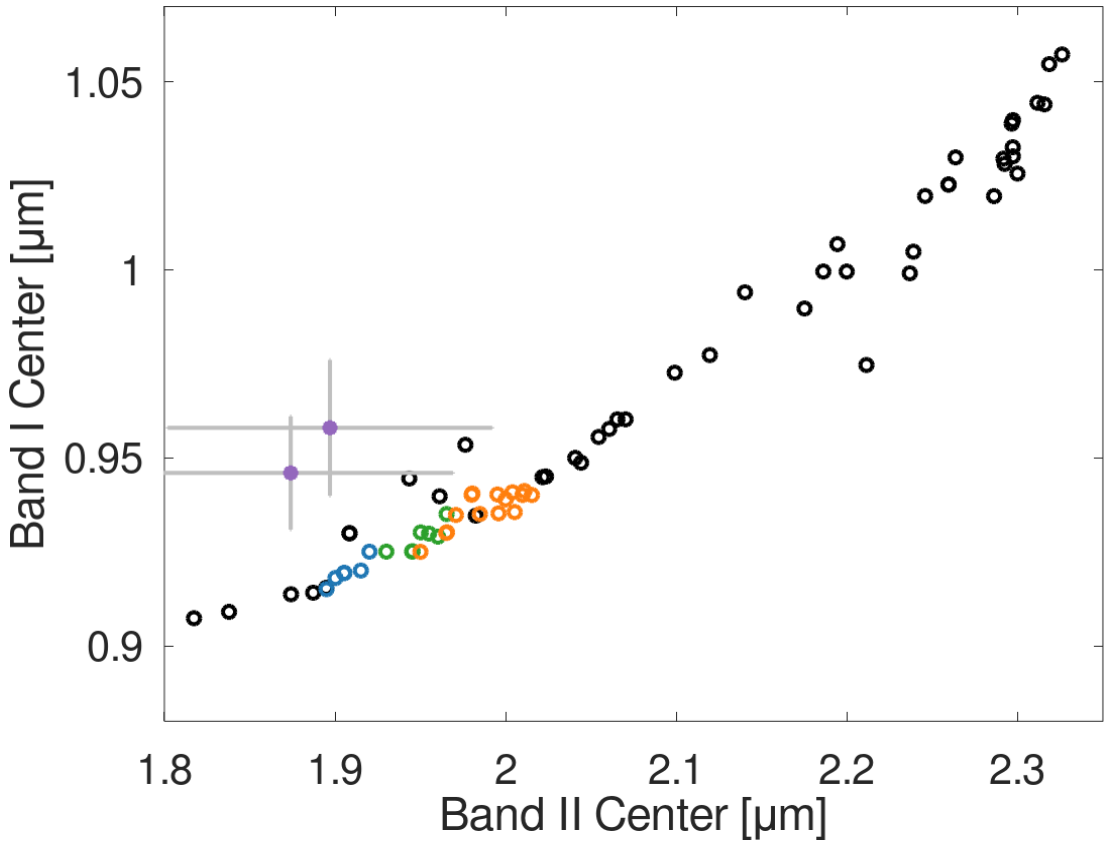

**Figure 14.** The band centers of the two boulders (violet dots) compared to those of synthetic pyroxenes (black circles) from Klima et al. (2008), and of diogenite (blue), howardite (green), eucrite (orange) meteorites (De Sanctis et al. 2013).

## 4. Discussion
In this section, we discuss the origin of the proposed exogenic materials found on Bennu. The timing of contamination by exogenic materials is unknown. We consider two possible hypothetical timings of contaminations: One is when the planetesimals and planets are dynamically active during giant planet migrations (Sec. 4.1), and the other is when planetesimals and planets are dynamically cold and collisions occur between objects sharing similar orbital elements (Sec. 4.2).

### 4.1. Formation of breccia-like and inclusion-like objects
Brecciated textures and inclusions are observed in many types of meteorites (e.g., Bischoff et al. 2006). Such meteorites provide important information about the impact and reassembly history of asteroids. The clasts in breccias or constituting inclusions can be fragments from different asteroids or different lithologies from the same parent body produced by energetic

Accepted manuscript for MNRAS                                                                                                                20 events such as impacts. The scale of proposed exogenic clasts > 0.2 m is much larger than the clast size observed in brecciated meteorites found on Earth, suggesting that relatively large impactors hit Bennu's parent body to make these large-scale clasts (DellaGiustina et al. 2021).

Clasts or inclusions must have undergone lithification with the material making up their host rocks. Bennu's host rocks are much more porous than analogous carbonaceous meteorites (Rozitis et al. 2020), meaning that the host rocks of inclusions did not necessarily experience strong compaction or melting as a result of impact. This suggests that inclusions were lithified by subsequent mild heating after the collision of an impactor without very high temperature or pressure. Possible heating sources include short-lived radiogenic species such as $^{26}$Al (Urey, 1955; Tachibana et al., 2006) or an ejecta blanket (Fernandes and Artemieva, 2012). The first possibility, heating by decay of $^{26}$Al, requires the impact and reassembly events to occur early. Cooling timing of planetesimals is highly dependent on size and composition and has been estimated as tens of millions of years for a 100-km-sized planetesimal (Henke et al. 2012; Neumann et al. 2014; Bland and Travis 2017). This timing is much earlier than Vesta family formation, 1 Gyr ago (Marzari et al. 1996). The second possibility, heating by ejecta blanket, requires that Bennu's dark lithology (typically observed as the host matrix in breccias) be sampled from the upper surface of the parent body. This could be distinguished by analysis of isotopic compositions of the solar-wind–implanted noble gases, which typically affect the surface to a meter-scale depth (Nagao et al. 2011), in the samples that will be returned by OSIRIS-REx.

**4.2. Recent collisional history**

Currently the main source for the basaltic material in the Solar System is considered to be the asteroid (4) Vesta. Vesta is the largest differentiated asteroid (~525 km in diameter) with a basaltic crust (McCord et al. 1970). The Vesta collisional family includes more than 15,000 known members (Nesvornỳ et al. 2015). The results found by the NASA Dawn mission to Vesta confirmed that this family is the result of two large cratering events. One is the Rheasilvia crater with diameter of 500 ± 25 km which is the young crater retention age of this basin indicates that it was formed about 1 Gyr ago. Another is the Veneneia crater, where crater counts suggest an age of 2.1 ± 0.2 Gyr, with diameters of 400 ± 25 km (Thomas et al. 1997; Schenk et al. 2012; Marchi et al. 2012; Jaumann et al. 2012).

By using the spectrophotometric data obtained by VISTA Hemisphere Survey, which is a panoramic wide field Infra-Red sky survey, Mansour et al. (2020) found that at least 80% of the ejected basaltic material from (4) Vesta is missing or is not yet detected because it is fragmented in sizes smaller than 1 km. Indeed, the all-sky spectrophotometric surveys showed the presence of a large number of basaltic candidates (associated with V (for Vesta) taxonomic type) over the entire inner main belt. Carvano et al. (2010) identified a total of 2818 V-type candidates based on the data obtained from the Sloan Digital Sky Survey with the optical filters $u$, $g$, $r$, $i$, $z$. Licandro et al. (2017) and Popescu et al. (2018) identified 798 basaltic candidates in the observations performed by the VISTA-VHS survey using the near-infrared filters $Y$, $J$, and $Ks$. Follow-up spectroscopic surveys confirmed their identification with a success rate of about 90% (Moskovitz et al. 2008; DeSanctis et al. 2011; Hardersen et al. 2014; Ieva et al. 2016 and references therein; Migliorini et al. 2017 and references therein; Medeiros et al. 2019).

It is well established that Bennu highly likely migrated into the near-Earth space from the inner main belt (Campins et al. 2010; Walsh et al. 2013; Bottke et al. 2015). In addition, Bennu is a retrograde rotator (Lauretta et al. 2019), which results in the Yarkovsky effect moving Bennu towards Sun (Chesley et al. 2014; Farnocchia et al. in press). From this observational fact, Bennu entered near-Earth space through the ν6 resonance. This scenario is supported by the occurrence of basaltic materials on Bennu. Over 95% of V-type asteroids are in the inner main belt (Mansour et al. 2020). Moreover, dynamically, some of the V-type candidates have orbital proper elements like the C-complex inner main belt families (Fig. 16), from which Bennu is considered to originate.

In contrast, there is little overlap with the (84) Klio, (163) Erigone, (313) Chaldaea, (329) Svea, and (623) Chimaera families in the orbital proper element, suggesting less possibility of collision between those families and V-type asteroids. Similarly, the young and small (302) Clarissa and (752) Sulamitis families are not a likely source for Bennu, either. Ultimately, only the Nysa–Polana–Eulalia family complex is a likely region where Bennu or its precursor prior to the catastrophic disruption could accrete V-type asteroidal material.

Figure 15 shows that it is more likely to find exogenic material on Bennu similar with the one of S-type, rather than the basaltic ones. The S-type asteroids dominate the inner main belt. The largest family showing olivine-pyroxene compositions (associated with the S-complex) is the (8) Flora family. It has more than 13,000 identified members and an estimated age of ~1 Gyr (Nesvornỳ et al. 2015). The approximate boundaries of this family in the proper elements space are $2.17 < a_p < 2.33$ au, $0.109 < e_p < 0.168$, and $2.4° < i_p < 6.9°$ (were $a_p$ is the proper semi-major axis, $e_p$ is the proper eccentricity, and $i_p$ is the proper inclination). The original Flora-family parent body is not considered to have been differentiated (Burbine et al. 2017). The asteroids belonging to this family have compositions similar to those of LL ordinary chondrites and are considered to be the parent bodies of these meteorites.

The second largest S-type asteroid family, with low inclination orbits in the inner main belt, is (20) Massalia. It has about 6500 identified members and an age of 150±50 Myr (Nesvornỳ et al. 2015). Another S-type swarm of asteroids is present in the Nysa–Polana complex and appears to be associated with the parent body (135) Hertha (Dykhuis and Greenberg 2015). The age of the family is inferred from the Yarkovsky dispersion of its members to be 300±60 Myr. Further dynamical modeling for contamination by ordinary chondrites is provided by Le Corre et al. (2021). They suggest that ordinary chondrites most likely contaminated would be prior to the parent body's disruption. Even if we assume impacts are more recent events after the Solar System stabilized, the impacts between the C-complex parent body and S- or V-type asteroids are plausible according to the orbital elements. However, in this case the breccia-like or inclusion-like morphologies with high-porosity bedrock are difficult. The current impact velocity in the inner main belt is ~ 7 km/s, and it produces considerable heating and pressure around the impact point. One possibility is that the impactors are trapped by penetrating a porous medium (Yasui et al. 2012). This case should be at the surface layers of the parent body. Similar to the discussion in Sec. 4.1, this can be clarified by noble gas analyses of the returned samples.

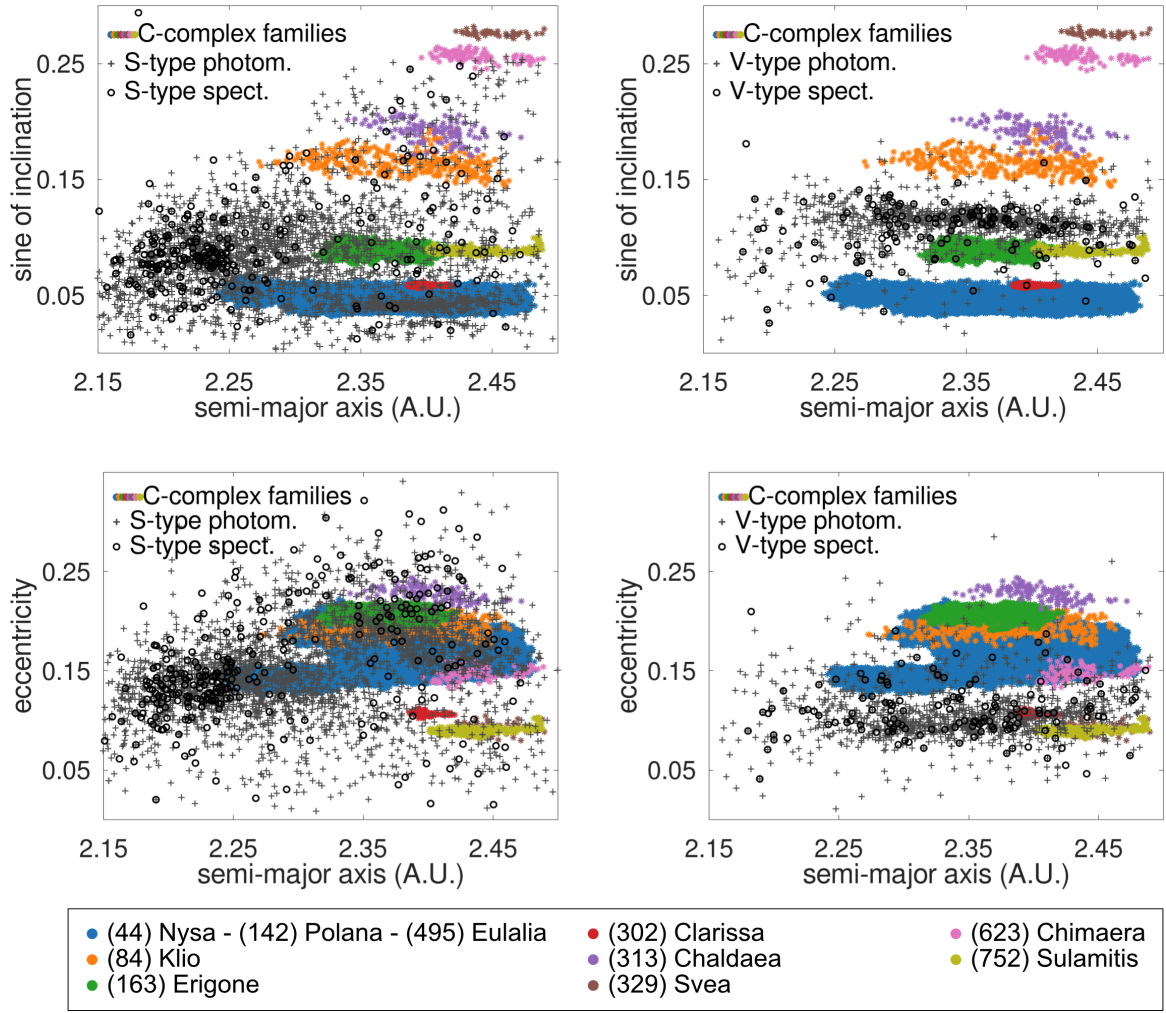

**Figure 15**. Proper semi-major axis versus eccentricity (top panel) and sine of proper inclination (bottom panel) for C-complex collisional families (blue dots) in the inner belt (de León et al. 2018), according to Nesvorný et al. (2015). We include photometry (grey crosses) and spectra (black dots) of asteroids with olivine-pyroxene compositions (S-types) and basaltic asteroids (V-types) for comparison. Photometric data were obtained from Carvano et al. (2010), Licandro et al. (2017), and Popescu et al. (2018), and spectra were retrieved from the Small Main-Belt Asteroid Spectroscopic Survey (SMASS) and Small Solar System Objects Spectroscopic Survey (S3OS2) surveys (Bus and Binzel 2002; Lazzaro et al. 2004), Vernazza et al. (2014), Ieva et al. (2016), Migliorini et al. (2017), and Medeiros et al. (2019).

**4.3. Comparison of exogenic materials on (101955) Bennu and (162173) Ryugu**

Bright exogenic boulders have also been found on Ryugu, another dark rubble-pile asteroid, which was the target of the sample return mission Hayabusa2 (Tatsumi et al. 2021). The telescopic Optical Navigation Camera (Kameda et al. 2017; Tatsumi et al. 2019) and the Near Infrared Spectrometer (Iwata et al., 2017) onboard the Hayabusa2 spacecraft observed six anhydrous-silicate–rich boulders. These boulders do not show strong 2-µm features, suggesting ordinary chondritic material (Tatsumi et al. 2021). Moreover, there could be two compositionally different groups of bright exogenic boulders on Ryugu according to the visible and near-infrared remote-sensing analysis: one with shallow 1-µm band absorption and another with deep absorption (Sugimoto et al. in press). Both Bennu and Ryugu show compositional diversity among their exogenic boulders, which may reflect that the mixture process with other types of asteroids is common in the catastrophic disruption cascade.

We compared spectra of the two largest exogenic boulders found on Ryugu, M7 and M13 (one each from the two groups in Tatsumi et al. (2020)) with our proposed exogenic materials on Bennu (Fig. 16). EX11 (which corresponds to Site 1 in DellaGiustina et al. 2021) has a much deeper absorption towards 1-µm than M7 and M13, which is consistent with the fact that EX11 matches HED meteorites showing an absorption around 2 µm (DellaGiustina et al. 2021). EX10 and EX29, showing blue slopes from v to x band, have also very different spectral shapes compared to M7 and M13. Their visible spectra are similar even though they have very different morphologies. EX36 has the closest spectrum to that of M13, although it presents slightly deeper absorptions near the UV and 1 µm. However, it should be noted that mixing of M13 with Bennu's average spectrum cannot reproduce the EX36 spectrum. Moreover, the range of reflectance factor is similar for EX36 and M13, meaning that they may have similar ordinary chondritic compositions. Both have breccia-like morphologies (see Section 3.3), likely mixtures of bright and dark compositions, suggesting that both asteroids have undergone impacts by multiple objects with different compositions.

Another important difference is that proposed exogenic material is much more abundant on Bennu than Ryugu. For sizes > 0.5 m, the number densities of exogenic objects on Ryugu and Bennu are ~1 km$^{-2}$ and ~40 km$^{-2}$, respectively, assuming that the surface area of Ryugu is 2.5 km$^2$ (Hirata et al. 2020) and the surface area of Bennu is 0.782 km$^2$ (Lauretta et al. 2019). Bennu presents much more contamination from exogenic materials, which likely reflects differences in impact conditions, such as impact velocity, angle, and size of impactor. These conditions could be informed by future collisional and dynamical evolution models of inner belt families.

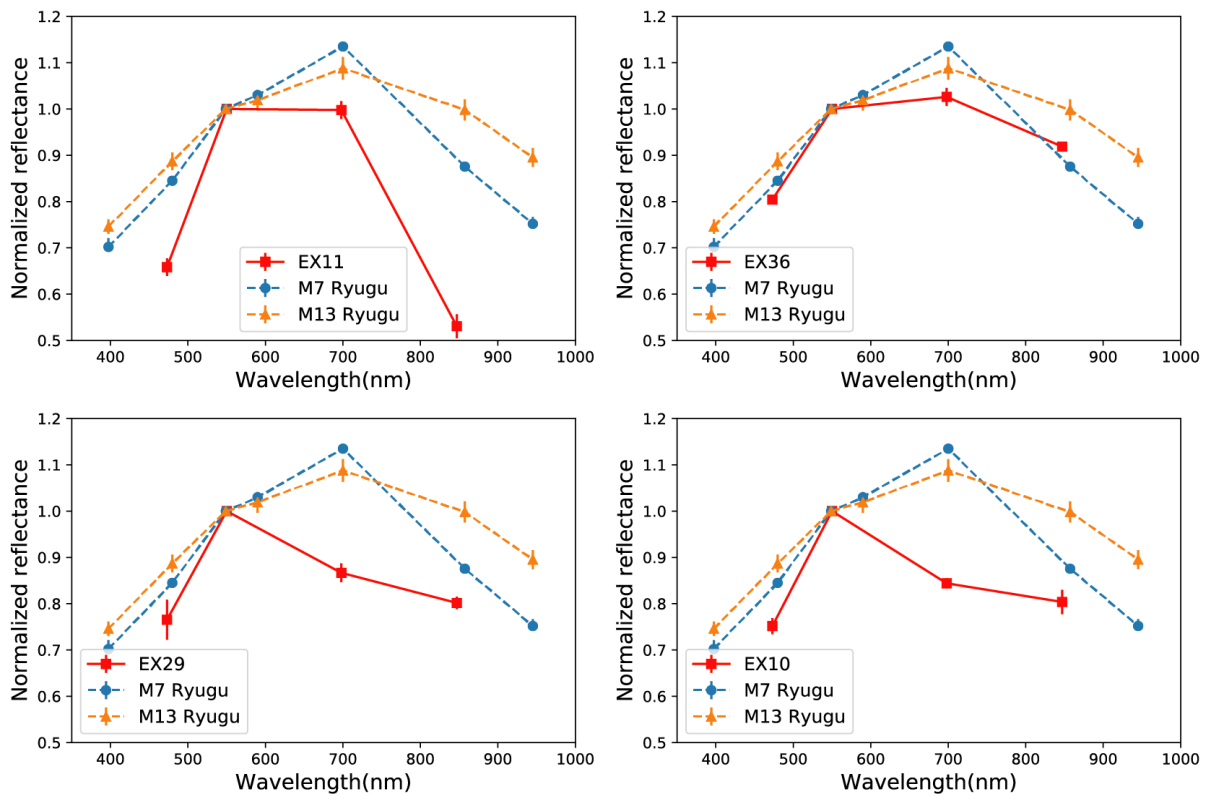

**Figure 16.** Visible spectral comparison between exogenic materials on Bennu (red) and exogenic bright boulders on Ryugu (blue and orange; Tatsumi et al. 2020).

## 5. Conclusions

We investigated materials on the surface of (101955) Bennu that show an x-band (0.847 micron) absorption indicative of anhydrous silicates, which we propose to be exogenic. Using the OSIRIS-REx MapCam imager with four bandpass filters, we found 77 instances of exogenic material widely distributed across Bennu's surface. We extracted visible reflectance spectrophotometry for 46 of these (>0.4 m) from the MapCam images and classified their morphological characteristics using higher-spatial-resolution images from the telescopic PolyCam. We also analyzed OVIRS near-infrared reflectance spectra of four of these instances that were obtained during close-range observations.

The visible reflectance spectra measured for 46 of the exogenic candidates are significantly brighter and have deeper near-UV absorptions than the typical boulders on Bennu (DellaGiustina et al. 2020), consistent with characteristics of anhydrous silicate–rich materials. PCA of MapCam spectra shows at least two major trends (Trend I and Trend II), suggesting mixing of Bennu's average composition with two endmembers: one with a deep 1-µm band absorption, possibly indicating pyroxene-rich material (Trend I), and the other with a shallow 1-µm absorption (Trend II). The ones with deeper 1-µm band absorptions, which make up Trend I, only match with HED meteorite spectra. Those with shallow 1-µm band absorptions, which make up Trend II, match spectra of either HED meteorites, ordinary chondrites, or carbonaceous chondrites. This conclusion is consistent with Le Corre et al. (2021), although we cannot rule out the possibility of HED and carbonaceous chondrites for an impactor material. The OVIRS near-infrared reflectance spectra are also consistent with the compositional diversity inferred from the colors, showing different absorption depths in the 1- and 2-µm bands.

The spectrophotometry of proposed exogenic objects with a spectral peak in the v-filter are matched particularly well by the Macibini Cl.3 melt sample. This is consistent with impact melt resulting from HED-like material colliding with Bennu's parent body, which could constrain impact models. It should be noted that the majority of the proposed exogenic materials are darker than HEDs or ordinary chondrites. Although the mixing may explain this lower albedo, laboratory spectra of some carbonaceous chondrites match some of our exogenic candidates in visible wavelength, especially REFF(30°,0°,30°) < 5%. Thus, we cannot exclude an endogenic origin for the relatively dark candidates in Trend II.

Morphologic expressions of the proposed exogenic materials include homogeneous rocks, breccia-like rocks, inclusion-like features, and others. The most common is inclusion-like, which is observed all over the surface. Inclusion-like features are usually observed in dark and cauliflower-like host rocks. The brightest candidates are preferentially homogeneous or breccia-like rocks.

The timing of contamination by these exogenic materials is unclear. Lithification of inclusion-like and breccia-like features might result from thermal processes, suggesting possible formation during the heat release from the short-lived radiogenic species, such as $^{26}$Al, which occurred very early in Solar System history. Alternatively, recent heating in ejecta blankets at the surface of the parent body is also plausible. On the other hand, more recent collisions of V-type asteroids, the most likely source of the basaltic meteorites, and S-type asteroids in inner main belt is also possible. Both S-type and V-type asteroids are widely distributed in the inner main belt and have similar orbital elements to C-complex families. V-type asteroids are concentrated in a particular region in inner main belt. We can conclude based on this distribution overlaps with C-comples families that the transition of Bennu to the near Earth through ν6 resonance, Nysa-Polana-Eulalia family is the most likely source of Bennu. This is consistent with the orbital calculations by Campins et al. (2010) and Walsh et al. (2013). We expect that the returned samples will reveal or constrain further the origin of those materials.

The comparison of proposed exogenic materials on Bennu with those on Ryugu suggests different compositions and abundances between the two asteroids, indicating different impact histories for these two bodies. Both asteroids show diversity in exogenic compositions on their surfaces, indicating they have undergone multiple impacts with objects of different compositions.

**Acknowledgments**

The authors thank Dr. De Sanctis for the careful and critical reviews. This material is based upon work supported by NASA under Contract NNM10AA11C issued through the New Frontiers Program. We are grateful to the entire OSIRIS-REx Team for making the encounter with Bennu possible. ET is supported by JSPS Core-to-Core program "International Network of Planetary Sciences". MP acknowledges a grant of the Romanian National Authority for Scientific Research - UEFISCDI, project number PN-III-P1-1.1-TE-2019-1504.

**Data Availability**

OCAMS MapCam and PolyCam data (Rizk et al. 2019) from the Detailed Survey–Baseball Diamond mission phase, and OVIRS data (Reuter et al. 2019) from the Recon A mission phase, are available via the Planetary Data System (https://sbn.psi.edu/pds/resource/orex/). Shape models of Bennu, including v28, are available via the Small Body Mapping Tool at http://sbmt.jhuapl.edu/.

**Appendix A. PolyCam images of the proposed exogenic objects**
All of the proposed exogenic materials detected in this study were resolved by PolyCam and are shown in Fig. A1 with different sizes of FOV.

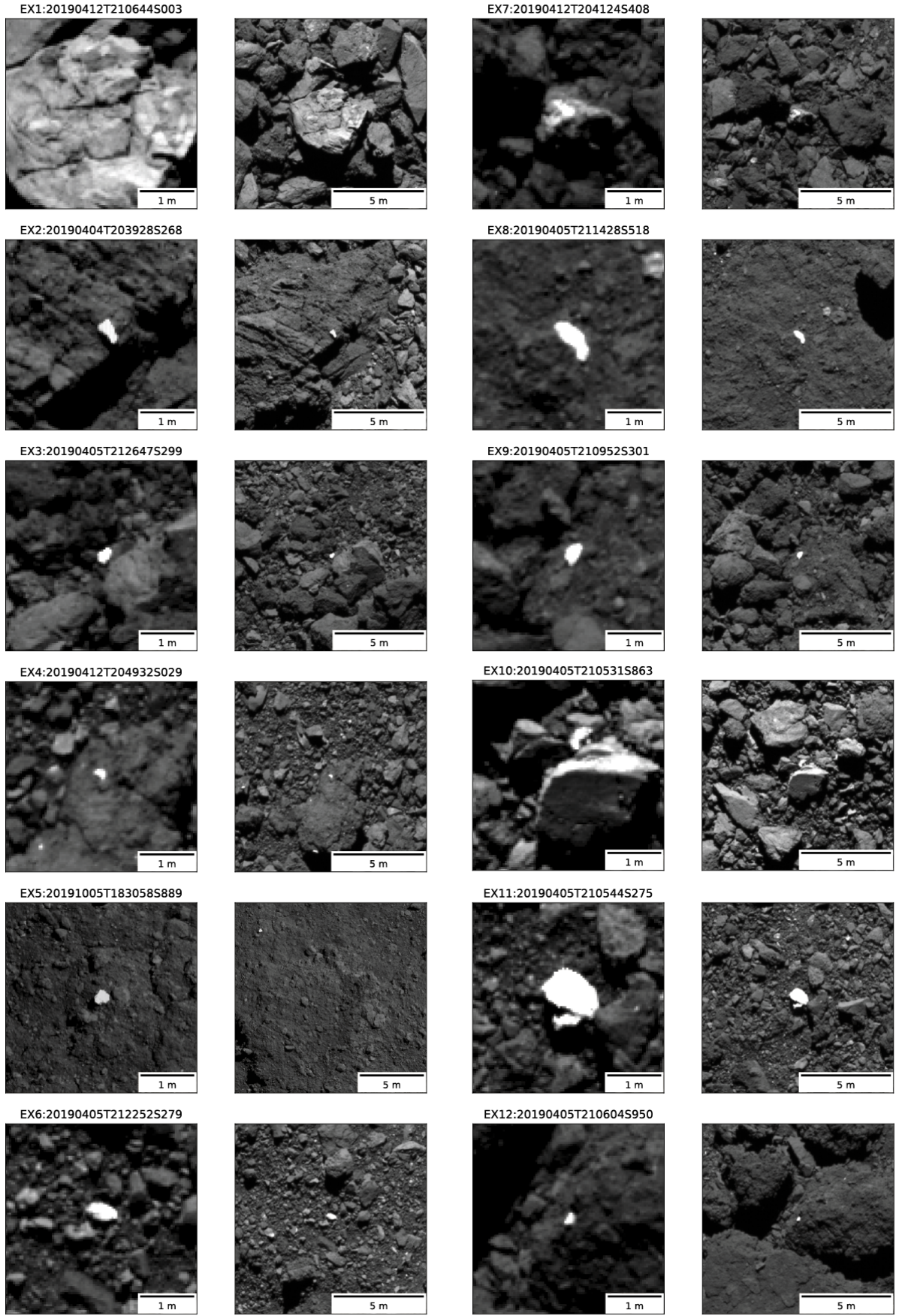

Figure A1 (continuous)

                                                                                                                                31

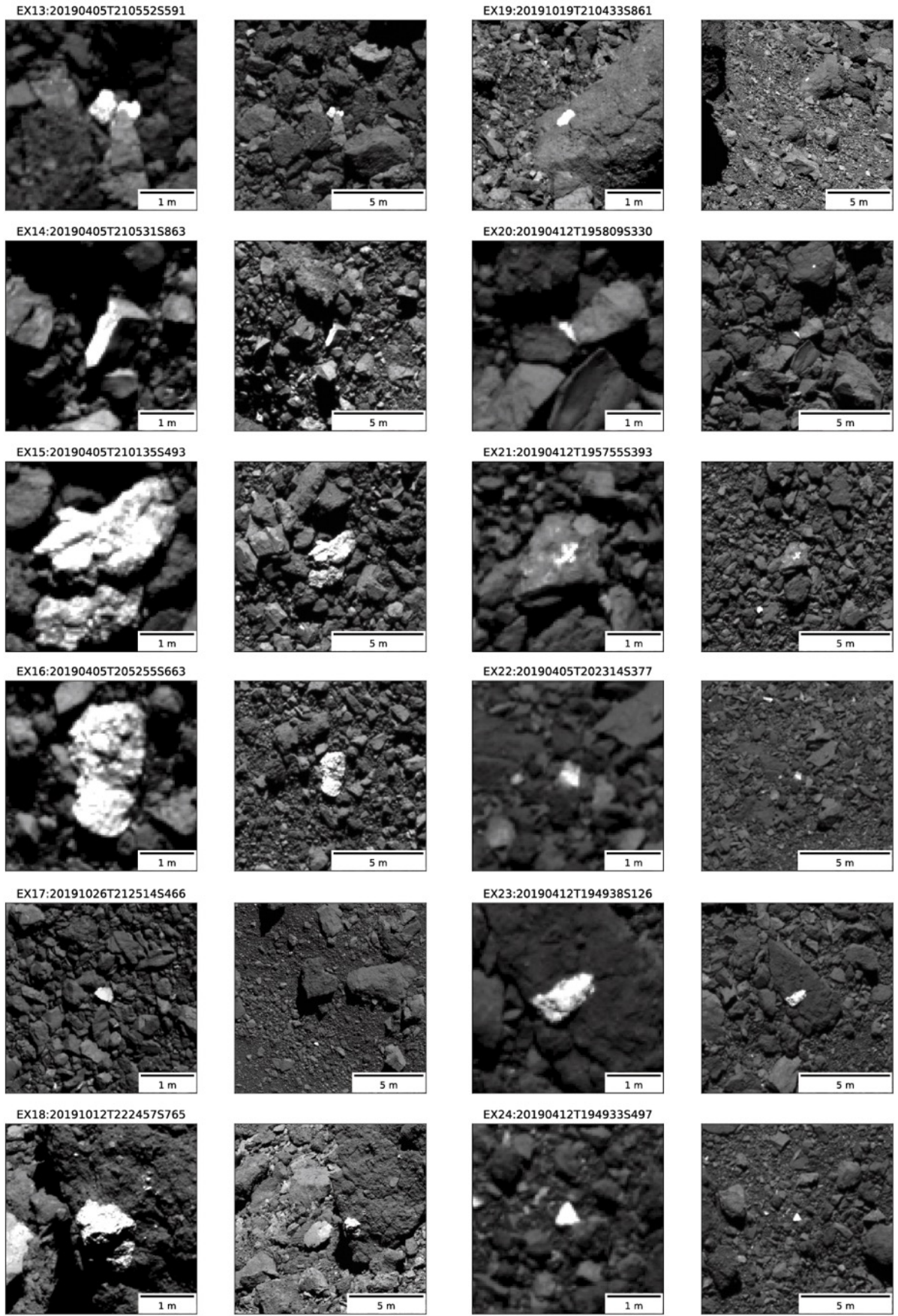

Figure A1 (continuous)

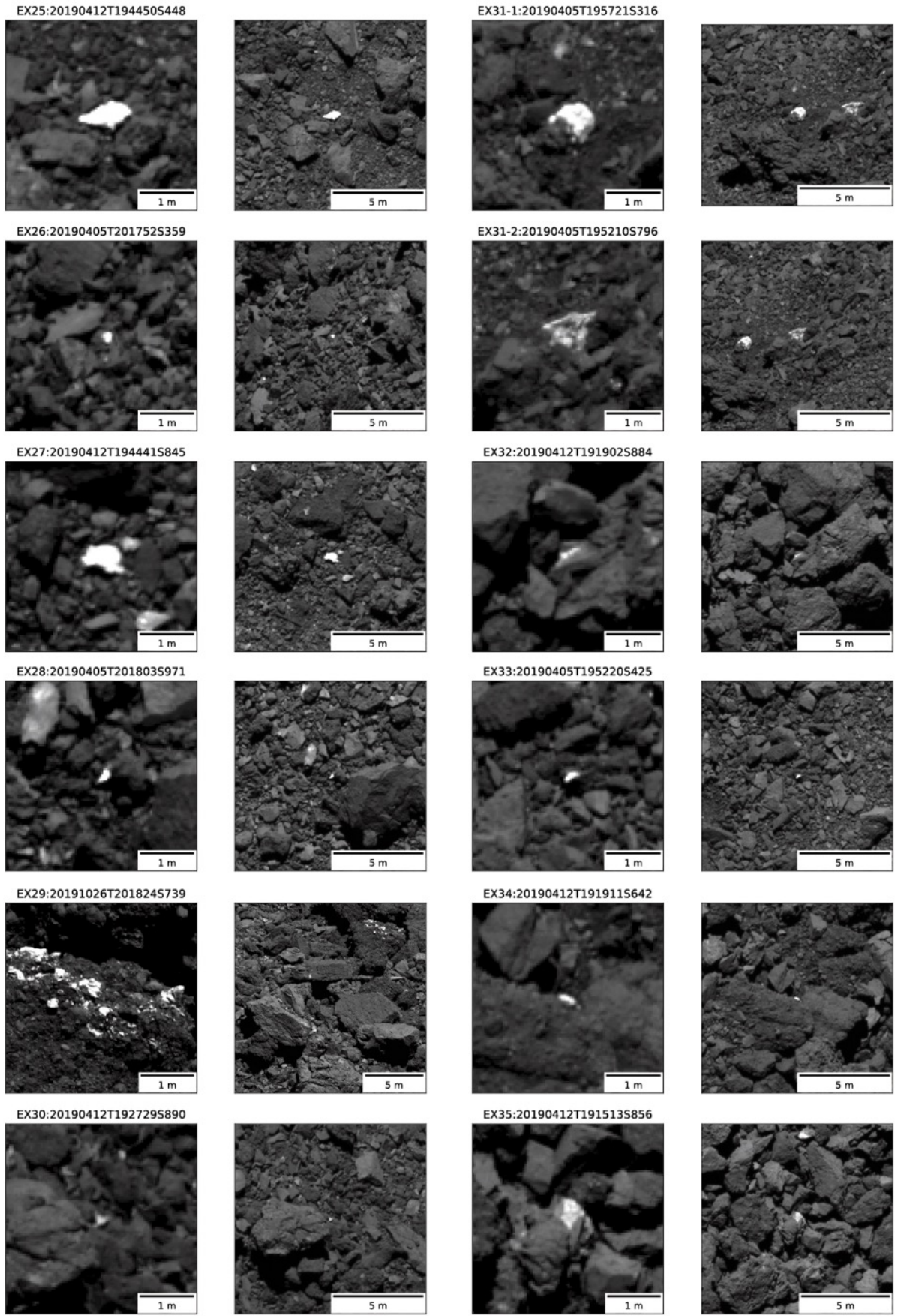

Figure A1 (continuous)

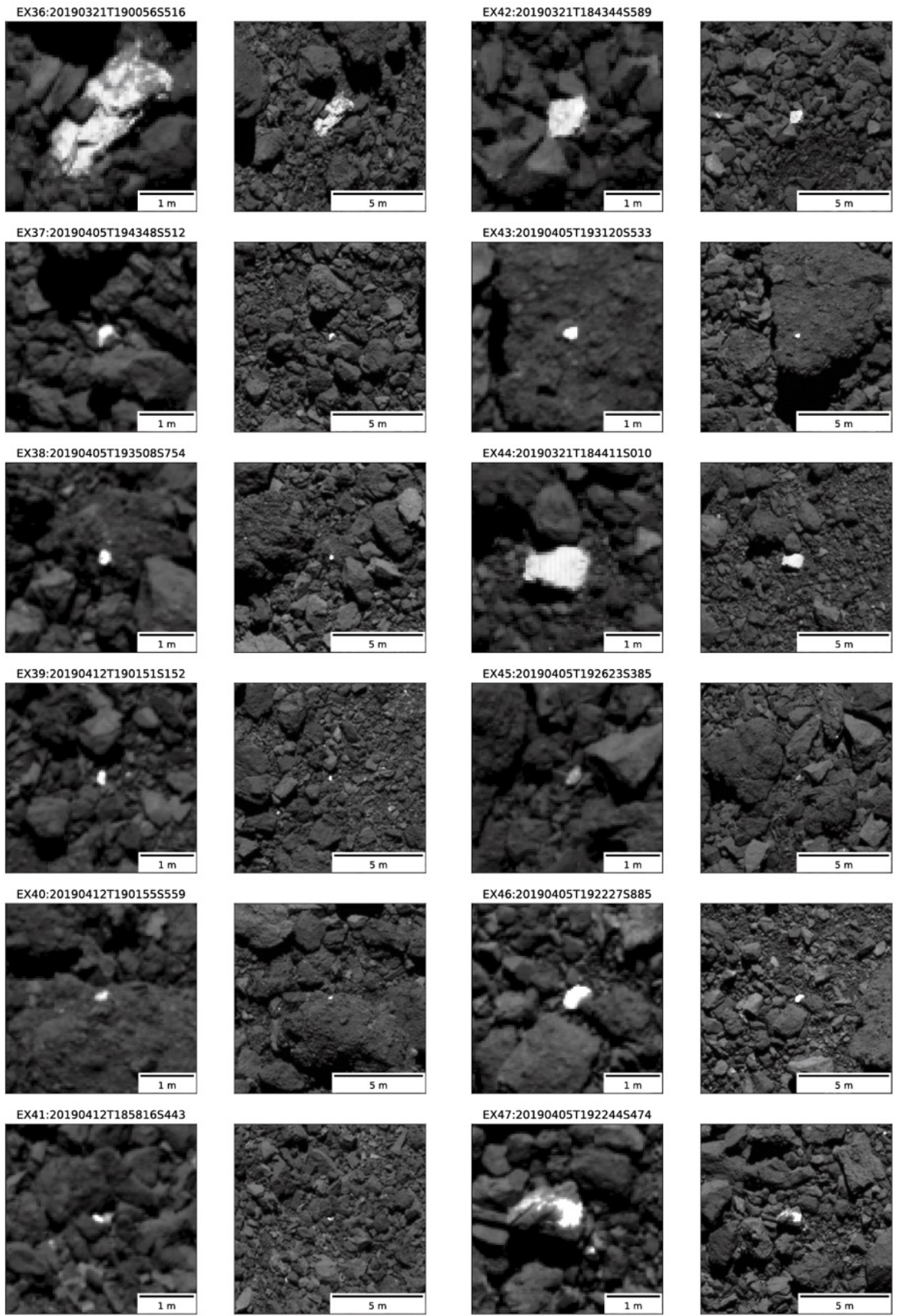

Figure A1 (continuous)

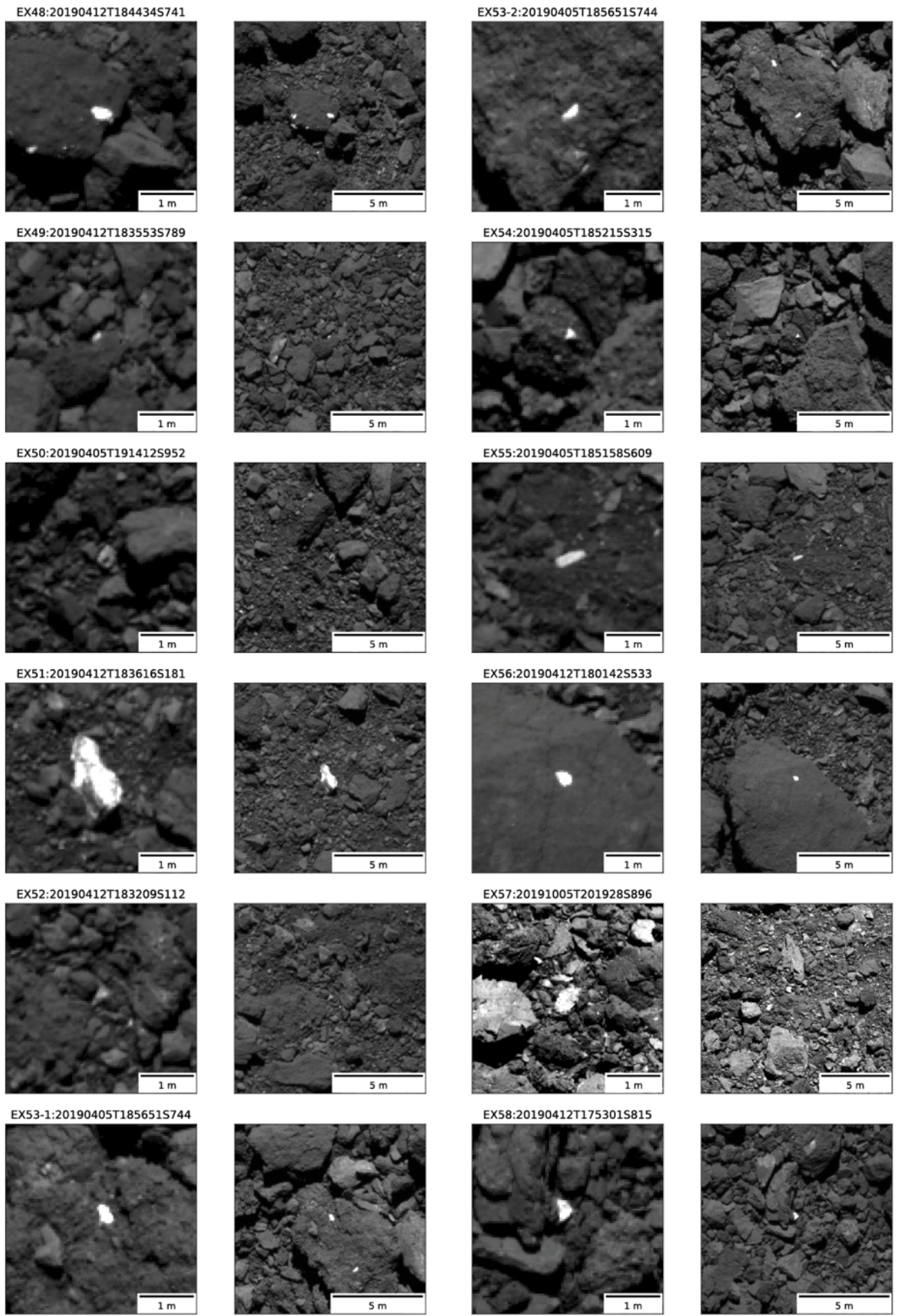

Figure A1 (continuous)

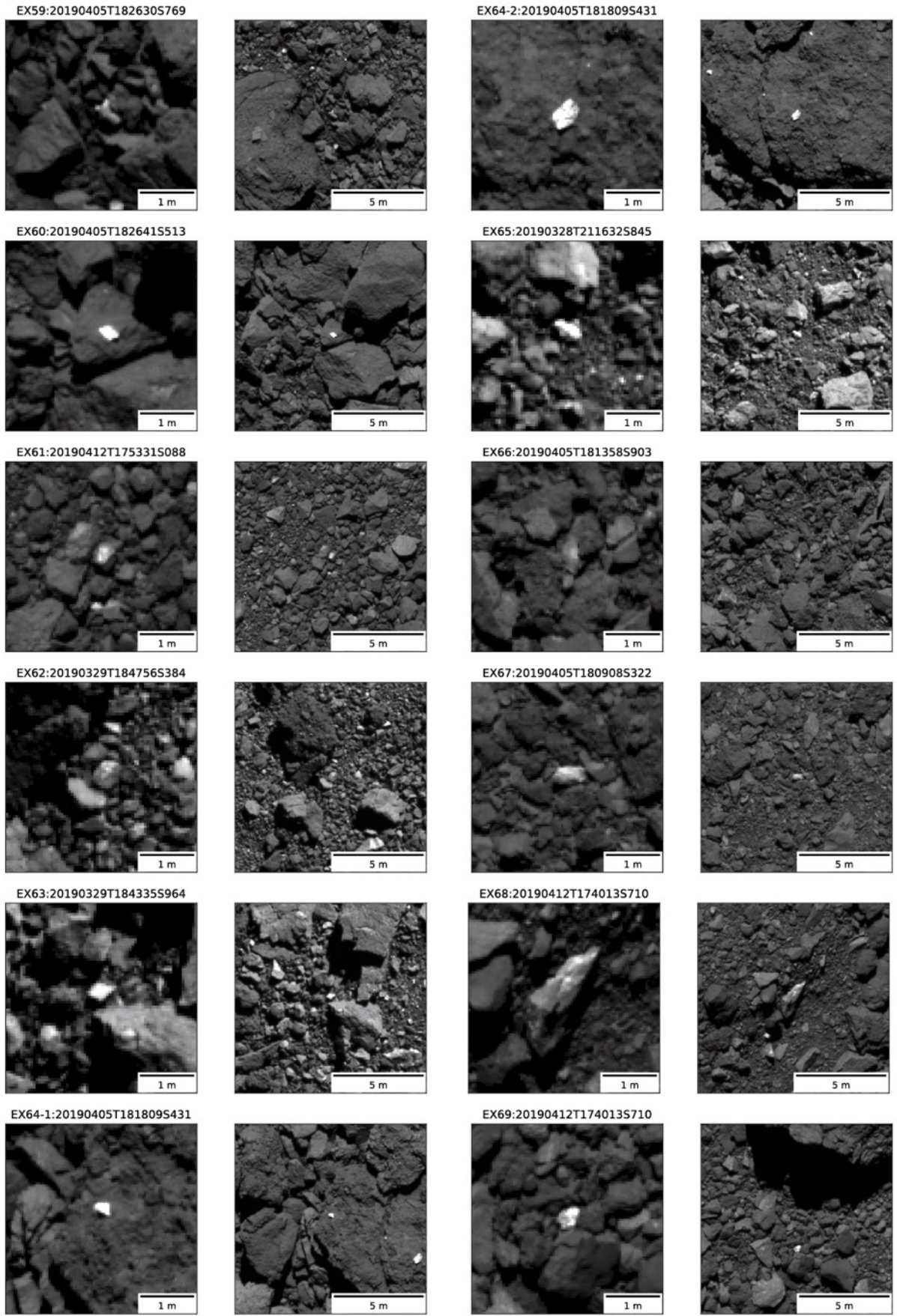

Figure A1 (continuous)

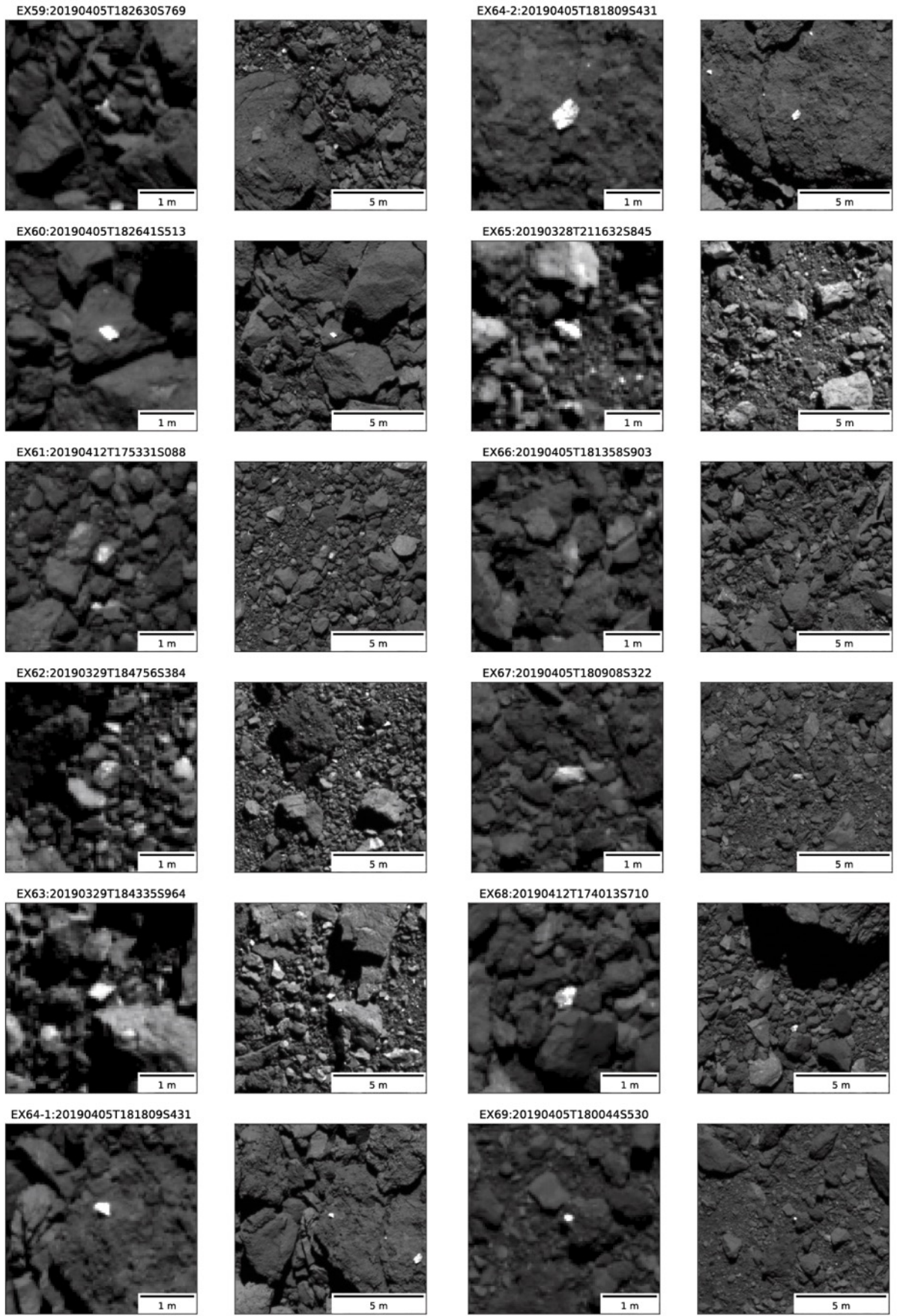

Figure A1 (continuous)

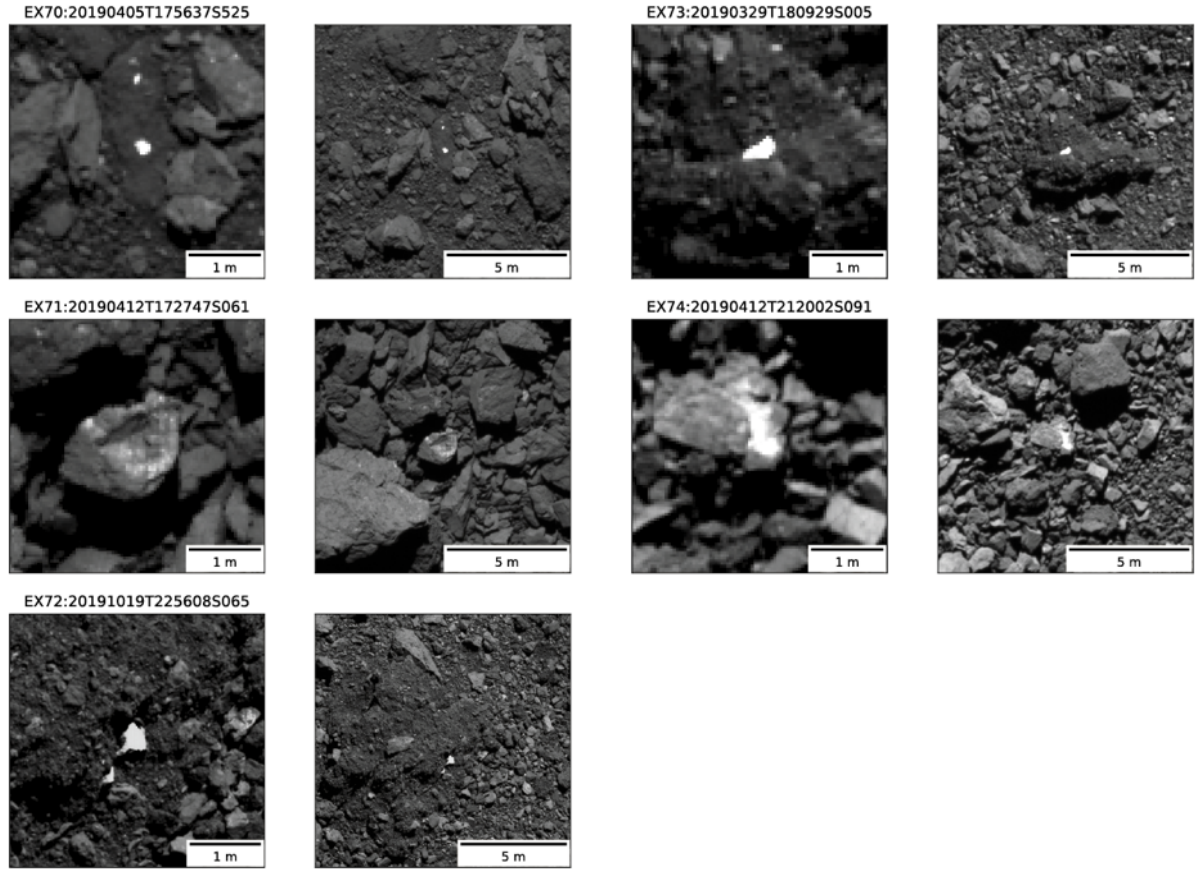

Figure A1 (continuous)

**Figure A1.** PolyCam images of the proposed exogenic objects. Each pair of images shows a close-up view. (first and third columns) and zoomed-out view (second and fourth columns) of the same feature. The exogenic object ID (Table 1) and time in UTC of image collection are shown above the close-up view. Images taken on 21 to 29 March 2019 have a pixel scale of 4.5 – 4.8 cm/pixel, images taken on 4 to 12 April 2019 have a pixel scale of 3.6–3.9 cm/pixel, and images taken on 5 to 26 October 2019 have a pixel scale of 1.2–1.6 cm/pixel.

## Appendix B. Comparison with specific meteorites

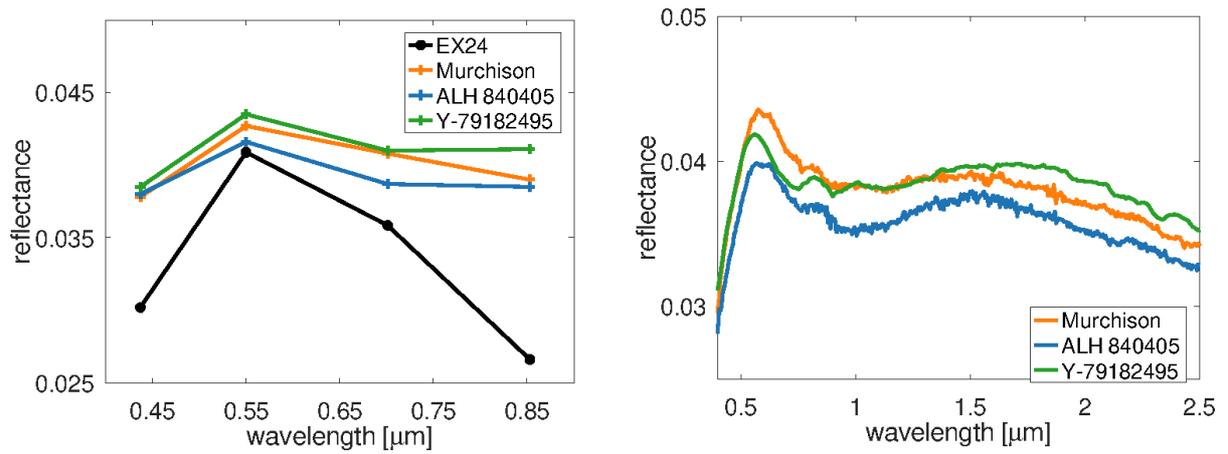

**Figure B1.** *Left*: MapCam spectrum of EX24 (black) compared with spectrophotometric data of three CM2 carbonaceous chondrites: a sample of separates from Murchison (SampleID: MS-CMP-002-B, FileIDs: CBMS02, orange curve), ALH 840405 (particulated to 125-250 μm, SampleID: MP-TXH-198-C, FileID: C1MP198C, blue curve) and Y-79182495 (SampleID: MP-TXH-031, FileID: C1MP31, shown in green). Right: The Relab visible-to-near-infrared spectra of the three meteorites (MS-CMP-002-B).

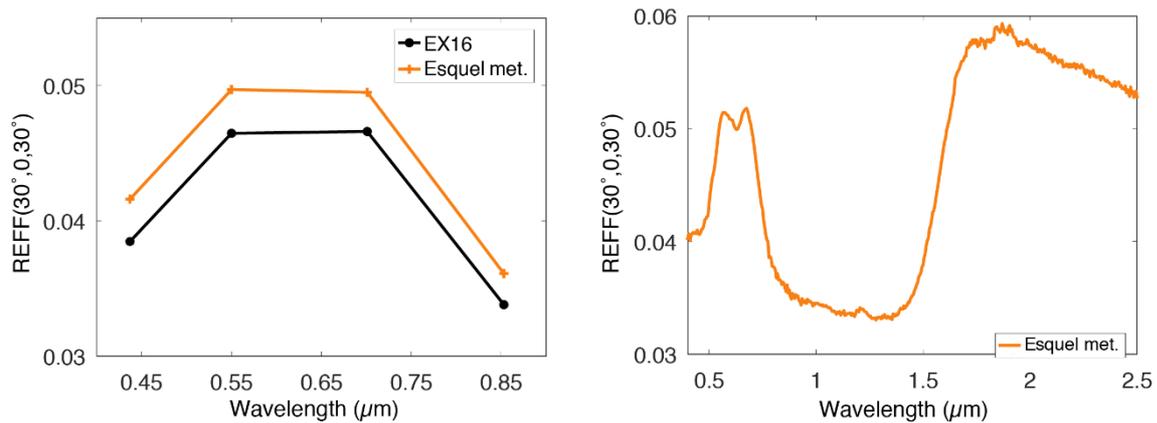

**Figure B2**. *Left*: MapCam spectrum of EX16 (black) *versus* the spectrophotometric data of the Esquel pallasite (SampleID: MB-TXH-043, FileID: C6MB43, in orange). *Right*: The visible-to-near-infrared spectrum of Esquel pallasite (MB-TXH-043).